\documentclass[a4paper,12pt]{article}

\pdfoutput=1
\usepackage{amsmath}
\usepackage{amssymb}
\usepackage{amsfonts}
\usepackage{mathrsfs}
\usepackage{bbm}
\usepackage{graphicx,subfigure}

\usepackage{bookmark}

\usepackage{cite}

\usepackage{calc}
\usepackage{hyperref}
\usepackage{array}

\usepackage{ulem}

\usepackage{multirow}
\usepackage{color}
\usepackage{xcolor,colortbl}

\usepackage{psfrag}
\usepackage{pstricks}
\usepackage{epsfig}

\allowdisplaybreaks

\newlength{\dinwidth}
\newlength{\dinmargin}
\definecolor{nicered}{rgb}{1.0,0.0,0.2}

\definecolor{color1}{rgb}{0.9,.4,.2}

\definecolor{color2}{rgb}{0.3,.6,.7}

\definecolor{color3}{rgb}{0.7,.2,.7}

\usepackage{color}

\begin{document}

\title{\bf Probing new physics in semileptonic  $\Sigma_b$ and $\Omega_b$ decays}

\author{ Jin-Huan Sheng$^{1}$\footnote{jinhuanphy@aynu.edu.cn}, Jie Zhu$^{1}$, Xiao-Nan Li$^{1}$,
 Quan-Yi Hu$^{1}$ and Ru-Min Wang$^{2}$\footnote{ruminwang@sina.com}  \\
{$^1$\small School of Physics and Electrical Engineering},\\
{\small Anyang Normal University, Anyang, Henan 455000, China}.\\
{$^2$\small  College of Physics and Communication Electronic}, \\
{\small Jiangxi Normal University, Nanchang, Jiangxi 330022, China}.
}


\date{}
\maketitle
\bigskip\bigskip
\maketitle
\vspace{-1.2cm}

\begin{abstract}
Recently, several hints of lepton non-universality have been observed in the semileptonic B meson decays in terms of both in the neutral current ($b\to s l \bar{l}$) and charged current ($b\to c l \bar{\nu_{l}}$) transitions.
Motivated by these inspiring results,  we perform the analysis of the  baryon decays  ${\Sigma_b}\to \Sigma_c l \bar{\nu_{l}}$ and $\Omega_b\to \Omega_c l \bar{\nu_{l}} (l=e,\mu,\tau)$ which are   mediated by $ b \to c l \bar{\nu_{l}}$ transitions at the  quark level, to scrutinize the nature of new physics (NP) in the model independent method.
We first use  the experimental measurements of  $\mathcal{B}( B\to D^{(*)} l\bar{\nu_{l}})$, ${\rm R}_{D^{(*)}}$  and ${\rm R}_{J/\psi}$ to constrain the NP coupling parameters in a variety of scenarios.
Using the constrained NP coupling parameters,
we report numerical results on various
observables related to the processes  ${\Sigma_b}\to \Sigma_c l \bar{\nu_{l}}$ and $\Omega_b\to \Omega_c l \bar{\nu_{l}}$, such as the branching ratios, the ratio of branching fractions,  the lepton side forward-backward asymmetries, the hadron and lepton longitudinal polarization asymmetries and the convexity parameter.
We also provide  the $q^2$ dependency of these observables  and
 we hope that the corresponding numerical results in this work will be testified by future experiments.
\end{abstract}

\newpage

\section{Introduction}
Though the Standard Model (SM) is considered as the most fundamental and  successful theory which describe almost all the phenomena of the particle physics, there are still some open issues that are not discussed in the SM, like matter-antimatter asymmetry, dark matter, etc.
Although there is no direct evidence for NP beyond the SM has been found, some possible hints of NP have been observed in the B meson decay  processes~\cite{Ray:2019gkv,Chang:2016cdi,Li:2018lxi,Bifani:2018zmi}.
Even though the SM gauge interactions are lepton flavor universal, the hints of lepton flavor universal violation (LFUV) have also been observed in several anomalies relative to the semileptonic B meson decays.
The most basic  experimental measurements which substantiate these anomalies are the ratio of the branching ratios ${\rm R}_{D^{(*)}}$ for
$b\to c l \bar{\nu}_l$ decay processes.
The ratio which is defined as ${\rm R}_{D^{(*)}}=\frac{\mathcal{B}(B\to D^{(*)}\tau \bar{\nu}_{\tau})}{\mathcal{B}(B\to D^{(*)}\ell \bar{\nu}_{\ell})}$
with $\ell=e,\mu$ has been measured first by the BaBar ~\cite{Lees:2012xj}.
Besides Belle and LHCb  also reported their results~\cite{Huschle:2015rga,Abdesselam:2016cgx,Abdesselam:2016xqt,Aaij:2015yra,Aaij:2017uff}.
The experimental measurement results for these anomalies show that there is  large deviations with their corresponding SM predictions.
Very recently, the Belle Collaborations announced the latest measurements of ${\rm R}_{D^{(*)}}$~\cite{Abdesselam:2019dgh}
\begin{eqnarray}
{\rm R}_D^{\rm Belle}&=&0.307\pm 0.037\pm 0.016, \nonumber \\
{\rm R}_{D^*}^{\rm Belle}&=&0.283\pm 0.017\pm 0.014,
\end{eqnarray}
which are in agreement with their SM predictions about within $0.2\sigma$ and $1.1\sigma$, respectively, and their combination agrees with the SM predictions within $1.2\sigma$.
Although the tension between the latest measurement results and their SM predictions is obviously reduced,
there is still $3.08\sigma$ corresponding SM predictions on combining all measurements in the global average fields.
The latest averaged results reported by Heavy Flavor Averaging Group (HFAG) are~\cite{HFAG}
\begin{eqnarray}\label{Eq.EXRD}
{\rm R}_D^{\rm avg}&=&0.340\pm 0.027\pm 0.013 , \nonumber \\
{\rm R}_{D^*}^{\rm avg}&=&0.295\pm 0.011\pm 0.008,
\end{eqnarray}
comparing with the SM predictions of ${\rm R}_{D^{(*)}}$~\cite{HFAG}
\begin{eqnarray}
{\rm R}_D^{\rm SM}=0.299\pm 0.003,~~~~ 
{\rm R}_{D^*}^{\rm SM}=0.258\pm 0.005.
\end{eqnarray}

One can see that above averaged experimental measurement results deviate from their SM predictions at $1.4\sigma$ and $2.5\sigma$ level, respectively.

Apart from ${\rm R}_D$ and ${\rm R}_{D^*}$ measurements, the ratio ${\rm R}_{J/\psi}$ has also been measured by LHCb~\cite{Aaij:2017tyk}
\begin{eqnarray}\label{Eq.EXRJP}
{\rm R}_{J/\psi}=\frac{\mathcal{B}(B_c\to J/\psi \tau \bar{\nu}_{\tau})}{\mathcal{B}(B_c\to J/\psi l \bar{\nu}_{l})}=0.71\pm 0.17\pm 0.18,
\end{eqnarray}
which central value prediction of the SM is in the range 0.25$\sim$0.28 and the experimental result has about $2\sigma$ tension with its SM prediction~\cite{Dutta:2017xmj,Wen-Fei:2013uea}.
The uncertainties arise from the choice of the approach for the $B_c \to J/\psi$ from factors~\cite{Wen-Fei:2013uea,Hu:2019qcn,Watanabe:2017mip,Wang:2018duy}.

These deviations  between the experimental measurements and their SM predictions
are perhaps from the uncertainties of hadronic transition form factors.
This may imply the lepton flavor universality is violated, which is the hint of the existence of NP.
Many works have been done based on  model independent framework ~\cite{Sakaki:2014sea,Bhattacharya:2016zcw,Feruglio:2018fxo,Jung:2018lfu,Hu:2018veh,Mu:2019bin,Huang:2018nnq}
or specific NP models by introducing new particles such as leptoquarks~\cite{Li:2016vvp,Yan:2019hpm,Barbieri:2016las},
SUSY particles~\cite{ Hu:2018lmk,Hu:2020yvs},
 charged Higgses ~\cite{Chen:2017eby,Iguro:2017ysu,Tanaka:2010se}, or new vector bosons~\cite{Matsuzaki:2017bpp}.

It is also  important and interesting to investigate the semileptonic baryon decays $\Sigma_b \to \Sigma_c l \bar{\nu}_l$ and $\Omega_b \to\Omega_c l \bar{\nu}_l$ which are mediated by the $b \to c l \bar{\nu}_l$ transition at the quark level.
Studying these processes not only can provide an independent determination of the Cabibbo-Kobayashi-Maskawa (CKM) matrix element $|V_{cb}|$, but also can confirm the LFUV in ${\rm R}_{\Sigma_c(\Omega_c)}$ which have a similar formalism to ${\rm R}_{D^{(*)}}$.
We will explore the NP effects on various observables for the $\Sigma_b \to \Sigma_c l \bar{\nu}_l$ and $\Omega_b \to\Omega_c l \bar{\nu}_l$ decays in the model independent effective field theory formalism.
It is necessary to study these decay modes both theoretically and experimentally to test the LFUV.
There will be several difficulties to measure the branching ratio $\mathcal{B}(\Sigma_b \to \Sigma_c l \bar{\nu}_l)$ because $\Sigma_b$ decay strongly and their branching ratios will be very small~\cite{Park:2005eka}.
Nevertheless it is feasible to measure $\mathcal{B}(\Omega_b\to \Omega_c l\bar{\nu}_l )$ as $\Omega_b$ decays predominantly weakly and the branching ratio is significantly large.
So it is worth to study these decay processes because they can provide  very comprehensive information about possible NP.

It will draw very interesting results to investigate the implications of ${\rm R}_{D^{(*)}}$ on the processes $\Omega_b\to \Omega_c l\bar{\nu}_l$ and $\Sigma_b \to \Sigma_c l \bar{\nu}_l$.
The authors of Refs.~\cite{Ebert:2006rp,Ivanov:1996fj,Ivanov:1998ya,Singleton:1990ye,Ke:2019smy,Ke:2012wa,Ke:2017eqo,Du:2011nj} give the total decay rate $\Gamma$(in units of $10^{10}s^{-1}$)  from $1.44$ to $4.3$ for $\Sigma_b \to \Sigma_c e \bar{\nu}_e$   and from 1.29 to 5.4 for $\Omega_b\to \Omega_c e\bar{\nu}_e$.
It is worthwhile to note that the complexity of the baryon structures and the lack of precise predictions of various form factors may lead to the variations in the prediction of the  total decay rate $\Gamma$.
In this paper we will give the predictions of various observables within SM and different NP scenarios.
Using the NP coupling parameters constrained from the latest experimental limits from $\mathcal{B}( B\to D^{(*)} l\bar{\nu_{l}})$, ${\rm R}_{D^{(*)}}$  and ${\rm R}_{J/\psi}$,
we investigate the NP effects of these anomalies on the differential branching fraction $d\mathcal{B}/dq^2$,  the ratios of branching fractions ${\rm R}_{\Omega_c(\Sigma_c)}(q^2)$, the lepton side forward-backward asymmetries $A_{FB}(q^2)$,
the longitudinal polarizations $P_L^{\Sigma_c(\Omega_c)}(q^2)$ of the daughter baryons $\Sigma_c(\Omega_c)$, the longitudinal polarizations $P_L^{l}(q^2)$ of the  lepton $l$  and the convexity parameter $C_F^l(q^2)$.
Note that there is different between our study and the Ref.~\cite{Rajeev:2019ktp}, in which
 $\Omega_b\to \Omega_c l\bar{\nu}_l$ and $\Sigma_b \to \Sigma_c l \bar{\nu}_l$ have also been investigated in a model independent way.
In our work the NP coupling parameters are assumed to be complex and
we consider the constraints on the NP coupling parameters from
the experimental limits of $\mathcal{B}( B\to D^{(*)} l\bar{\nu_{l}})$, ${\rm R}_{J/\psi}$ and ${\rm R}_{D^{(*)}}$.
However, NP coupling parameters are set to real  and only   ${\rm R}_{D^{(*)}}$ is considered in Ref.~\cite{Rajeev:2019ktp}.


Our paper is organized as follows. In Sec.\ref{Sec.two} we briefly
introduce the effective  theory describing the $b \to c l\bar{\nu}_l$ transitions as well as the form factors, the helicity amplitudes and some observables of the processes $\Omega_b\to \Omega_c l\bar{\nu}_l$ and $\Sigma_b \to \Sigma_c l \bar{\nu}_l$.
Sec.~\ref{Sec.result} is devoted to the numerical results and discussions for the predictions within the SM and  various NP scenarios. Our conclusions are given in Sec.~\ref{Sec.four}.


\section{Theory framework }\label{Sec.two}
The most general effective Lagrangian including both the SM and the NP contribution for $B_1\to B_2 l \bar{\nu}_l$ decay processes, where $B_1=\Sigma_b(\Omega_b)$, $B_2=\Sigma_c(\Omega_c)$, mediated by the quark level transition $b \to c l \bar{\nu}_l$ is given by \cite{Cirigliano:2009wk,Bhattacharya:2011qm}
{\small
\begin{eqnarray}\label{Eq.Lbtoclnul}
\mathcal L_{\rm eff} &=&
-\frac{4G_F}{\sqrt{2}}V_{c b}\,\Bigg\{(1 + V_L)\,\bar{l}_L\,\gamma_{\mu}\,\nu_L\,\bar{q}_L\,\gamma^{\mu}\,b_L \nonumber\\
&&+V_R\,\bar{l}_L\,\gamma_{\mu}\,\nu_L\,\bar{q}_R\,\gamma^{\mu}\,b_R \ +S_L\,\bar{l}_R\,\nu_L\,\bar{q}_R\,b_L  \nonumber\\
&&+S_R\,\bar{l}_R\,\nu_L\,\bar{q}_L\,b_R +
T_L\,\bar{l}_R\,\sigma_{\mu\nu}\,\nu_L\,\bar{q}_R\,\sigma^{\mu\nu}\,b_L  \Bigg\} + {\rm h.c.}\,,
\end{eqnarray}}
where $G_F$ is the Fermi constant, $V_{cb}$ is the CKM matrix elements and $(q,b,l,\nu)_{L,R}=P_{L,R}(q,b,l,\nu)$ are the chiral quark (lepton) fields with $P_{L,R}=(1\mp \gamma_5)/2$ as the projection operators.
Here we note that the NP coupling parameters $V_{L,R}$, $S_{L,R}$, $T_L $ characterizing the NP contributions coming from  the new  vector, scalar and tensor  interactions are associated with left handed neutrino and  these NP coupling parameters are all zero in the SM.
In our work  we  focus on a study of the vector and scalar type interactions, excepting the tensor interaction, and we assume that
the NP coupling parameters  $V_{L,R}$ and $S_{L,R}$ are complex.

\subsection{Form factors and helicity amplitudes}\label{FFandHA}

The hadronic matrix elements of vector and axial vector
currents for the decays $B_1\to B_2 l\bar{\nu}_l$  are parametrized in
terms of  various hadronic  form factors as follows:
{\small\begin{eqnarray}\label{Eq.FFVA}
	M^V_{\mu} &=& \langle B_2,\lambda_2\vert \bar{c}\gamma_{\mu}b\vert B_1,\lambda_1\rangle \nonumber\\&=& \bar{u}_2(p_2,\lambda_2)\big{[}f_1(q^2)\gamma_{\mu}+if_2(q^2)\sigma_{\mu\nu}q^{\nu}+f_3(q^2)q_{\mu}\big{]}\nonumber \\ &\times&
u_1(p_1,\lambda_1),\nonumber\\
	M^A_{\mu} &=& \langle B_2,\lambda_2\vert \bar{c}\gamma_{\mu}\gamma_5 b\vert B_1,\lambda_1\rangle \nonumber\\&=& \bar{u}_2(p_2,\lambda_2)\big{[}g_1(q^2)\gamma_{\mu}+ig_2(q^2)\sigma_{\mu\nu}q^{\nu}+g_3(q^2)q_{\mu}\big{]} \gamma_5 \nonumber \\ &\times&
 u_1(p_1,\lambda_1),\nonumber
	\end{eqnarray}
}
where $\sigma_{\mu\nu}=\frac{i}{2}(\gamma_{\mu}\gamma_{\nu}-\gamma_{\nu}\gamma_{\mu})$, ~$q_{\mu}=(p_1-p_2)_{\mu}$ is the four momentum transfer.
$\lambda_1$ and $\lambda_2$ are the helicities of the parent baryon $B_1$ and daughter
baryon $B_2$, respectively.
~Here $B_1$ represents the bottomed baryon $\Sigma_b$ or $\Omega_b$
and $B_2$ represents the charmed baryon $\Sigma_c$ or $\Omega_c$.
Using the equation of motion, we can obtain
the hadronic matrix elements of the scalar and pseudo-scalar currents between these two baryons.
The expressions for them can be written
{\footnotesize
\begin{eqnarray}
\langle B_2,\lambda_2\vert \bar{c}b\vert B_1,\lambda_1\rangle &=& \bar{u}_2(p_2,\lambda_2)\nonumber \\ &\times& \left[f_1(q^2)\frac{q}{m_b-m_c}+f_3(q^2)\frac{q^2}{m_b-m_c}\right]\nonumber \\ &\times& u_1(p_1,\lambda_1),\nonumber\\
\langle B_2,\lambda_2\vert \bar{c}\gamma_5 b\vert B_1,\lambda_1\rangle& =& \bar{u}_2(p_2,\lambda_2)\nonumber \\ &\times& \left[-g_1(q^2)\frac{q}{m_b+m_c}-g_3(q^2)\frac{q^2}{m_b+m_c}\right]\gamma_5 \nonumber \\ &\times&u_1(p_1,\lambda_1),\nonumber
\end{eqnarray}
}
where $m_b$ and $m_c$ are the respective masses of $b$ and $c$ quarks calculated at the renormalization scale $\mu =m_b$.

When both baryons are
heavy, it is also convenient to parametrize the matrix element in  the heavy quark limit, these matrix elements can be parametrized in terms of
four velocities $v^{\mu}$ and $v'^{\mu}$ as follows
{\small
\begin{eqnarray}
M^V_{\mu} &=& \langle B_2,\lambda_2\vert \bar{c}\gamma_{\mu}b\vert B_1,\lambda_1\rangle \nonumber\\&=& \bar{u}_2(p_2,\lambda_2)\left[F_1(w)\gamma_{\mu}+F_2(w)v_{\mu}+F_3(w)v'_{\mu}\right]u_1(p_1,\lambda_1),\nonumber\\
M^A_{\mu} &=& \langle B_2,\lambda_2\vert \bar{c}\gamma_{\mu}\gamma_5 b\vert B_1,\lambda_1\rangle \nonumber\\&=& \bar{u}_2(p_2,\lambda_2)\left[G_1(w)\gamma_{\mu}+G_2(w)v_{\mu}+G_3(w)v'^{\mu}\right]\gamma_5 \nonumber \\ &\times& u_1(p_1,\lambda_1),\nonumber
\end{eqnarray}
}
where $w=v.v'=\left(M^2_{B_1}+M^2_{B_2}-q^2\right)/2M_{B_1}M_{B_2}$,  $M_{B_1}$ and $M_{B_2}$ are the masses of the $B_1$ and $B_2$ baryons, respectively.
 The relationship of these two sets of form factors are related via~\cite{Dutta:2018zqp}
\begin{eqnarray}
f_1(q^2)& =& F_1(q^2)+\left(m_{B_1}+m_{B_2}\right)\left[\frac{F_2(q^2)}{2m_{B_1}}+\frac{F_3(q^2)}{2m_{B_2}}\right],\nonumber\\
f_2(q^2) &=& \frac{F_2(q^2)}{2m_{B_1}}+\frac{F_3(q^2)}{2m_{B_2}},\nonumber\\
f_3(q^2) &=& \frac{F_2(q^2)}{2m_{B_1}}-\frac{F_3(q^2)}{2m_{B_2}},\nonumber\\
g_1(q^2)& =& G_1(q^2)-\left(m_{B_1}-m_{B_2}\right)\left[\frac{G_2(q^2)}{2m_{B_1}}+\frac{G_3(q^2)}{2m_{B_2}}\right],\nonumber\\
g_2(q^2) &=& \frac{G_2(q^2)}{2m_{B_1}}+\frac{G_3(q^2)}{2m_{B_2}},\nonumber\\
g_3(q^2) &=& \frac{G_2(q^2)}{2m_{B_1}}-\frac{G_3(q^2)}{2m_{B_2}}.
\end{eqnarray}

In our numerical analysis, we follow Ref.~\cite{Ebert:2006rp} and use the form factor inputs obtained in the framework of the relativistic quark model.
In the heavy quark limit, the form factors can be expressed in terms of the Isgur-Wise function $\zeta_1(w)$ as follows ~\cite{Ebert:2006rp,Ke:2012wa}
\begin{eqnarray}
F_1(w)& =& G_1(w) = -\frac{1}{3}\zeta_1(w),\nonumber\\
F_2(w)& =& F_3(w) = \frac{2}{3}\frac{2}{w+1}\zeta_1(w),\nonumber\\
G_2(w)& =& G_3(w) = 0,
\end{eqnarray}
and the values of $\zeta_1(w)$ in the whole kinematic range, pertinent for our analysis, were mainly obtained from Ref.~\cite{Ebert:2006rp}.

The helicity amplitudes can be defined by \cite{Gutsche:2015mxa,Shivashankara:2015cta,Ray:2018hrx,Zhang:2019xdm,Dutta:2018zqp}
\begin{eqnarray}
H^{V/A}_{\lambda_2,\lambda_W}&=&M^{V/A}_{\mu}(\lambda_2)\varepsilon^{\dagger\mu}(\lambda_W),
\end{eqnarray}
where $\lambda_2$ and $\lambda_W$ denote the respective helicities of the daughter baryon and $W^-_{off-shell}$,
In the rest frame of the parent baryon $B_1$,
the vector and axial vector hadronic helicity amplitudes in the terms of the various form factors and  NP coupling parameters are given by~\cite{Rajeev:2019ktp,Gutsche:2015mxa,Shivashankara:2015cta,Ray:2018hrx,Zhang:2019xdm,Dutta:2018zqp}

{\footnotesize
\begin{eqnarray}\label{Eq.HVA}
H_{\frac{1}{2}\,0}^V &=& (1+V_L+V_R )\frac{\sqrt{Q_-}}{\sqrt{q^2}}
\Big[(M_{B_1} + M_{B_2})\,f_1(q^2) - q^2\,f_2(q^2)\Big], \nonumber \\
H_{\frac{1}{2}\,0}^A &=& (1+V_L-V_R)\frac{\sqrt{Q_+}}{\sqrt{q^2}}
\Big[(M_{B_1} - M_{B_2})\,g_1(q^2) + q^2\,g_2(q^2)\Big], \nonumber \\
H_{\frac{1}{2}\,+}^V &=& (1+V_L+V_R)\sqrt{2\,Q_-}\,\Big[-f_1(q^2) + (M_{B_1} + M_{B_2})\,f_2(q^2)\Big], \nonumber \\
H_{\frac{1}{2}\,+}^A &=& (1+V_L-V_R)\sqrt{2\,Q_+}\,\Big[-g_1(q^2) - (M_{B_1} - M_{B_2})\,g_2(q^2)\Big], \nonumber \\
H_{\frac{1}{2}\,t}^V &=& (1+V_L+V_R)\frac{\sqrt{Q_+}}{\sqrt{q^2}}\,\Big[(M_{B_1} - M_{B_2})\,f_1(q^2) + q^2\,f_3(q^2)\Big], \nonumber \\
H_{\frac{1}{2}\,t}^A &=& (1+V_L-V_R)\frac{\sqrt{Q_-}}{\sqrt{q^2}}\,\Big[(M_{B_1} + M_{B_2})\,g_1(q^2) - q^2\,g_3(q^2)\Big], \nonumber
\end{eqnarray}
}
where $Q_{\pm}=(M_{B_1}\pm M_{B_2})^2-q^2$ and $f_i$,~$g_i$ ($i=1,2,3$) are the various form factors.
Either from parity or from explicit calculation, it is clear to find that $H^V_{-\lambda_2-\lambda_W}= H^V_{\lambda_2\lambda_W}$ and $H^A_{-\lambda_2-\lambda_W}= -H^A_{\lambda_2\lambda_W}$.
So the total left-handed helicity amplitude is
\begin{equation}
H_{\lambda_2\lambda_W} = H^V_{\lambda_2\lambda_W}-H^A_{\lambda_2\lambda_W}
\end{equation}

Similarly, the scalar and pseudoscalar helicity amplitudes associated with the form factors and NP coupling parameters $G_S$ and $G_P$ can be written as
{\footnotesize
\begin{eqnarray}
H_{\lambda_2\,0}^{SP} &=&H_{\lambda_2\,0}^{S}-H_{\lambda_2\,0}^{P},\nonumber \\
H_{\frac{1}{2}\,0}^{S} &=& \left(S_L+S_R\right)\,\frac{\sqrt{Q_+}}{m_b - m_{q}}\,\Big[(M_{B_1} - M_{B_2})\,f_1(q^2) + q^2\,f_3(q^2)\Big]\,, \nonumber \\
H_{\frac{1}{2}\,0}^{P} &=&\left(S_L-S_R\right)\,\frac{\sqrt{Q_-}}{m_b + m_{q}}\,\Big[(M_{B_1} + M_{B_2})\,g_1(q^2) - q^2\,g_3(q^2)\Big],\nonumber \label{Eq.HSP}
\end{eqnarray}
}
one can  see that
$H^S_{-\lambda_2-\lambda_W}= H^S_{\lambda_2\lambda_W}$ and $H^P_{-\lambda_2-\lambda_W}= -H^P_{\lambda_2\lambda_W}$.
The results of above helicity amplitudes in SM can be obtained by setting $V_{L,R}=0$ and $S_{L,R}=0$.

\subsection{The observables for $\Sigma_b \to \Sigma_c l \bar{\nu}_l$ and $\Omega_b \to \Omega_c l \bar{\nu}_l$}

After including the NP contributions, the differential decay distribution for $\Sigma_b \to \Sigma_c l \bar{\nu}_l$ and $\Omega_b \to \Omega_c l \bar{\nu}_l$
in term of  $q^2$, $\theta_l $  and helicity amplitudes can be written as~\cite{Shivashankara:2015cta,Dutta:2018zqp}
{\small
\begin{eqnarray}\label{Eq.dTdsdscos}
\frac{d^2\Gamma(B_1\to  B_2 l \bar{\nu}_l)}{dq^2dcos\theta_l}&=&N \left(1-\frac{m_l^2}{q^2}\right)^2\bigg{[}A_1+\frac{m_l^2}{q^2}A_{2}\nonumber \\
&& +2A_3 +\frac{4m_l}{\sqrt{q^2}} A_4\bigg{]},
\end{eqnarray}
}
where
{\small
\begin{eqnarray}
N&=&\frac{G_F^2 |V_{c b}|^2 q^2 \sqrt{\lambda(M_{B_1}^2, M_{B_2}^2, q^2)}}{2^{10} \pi^3 M_{B_1}^3},\nonumber\\
\lambda(a,b,c)& =& a^2+b^2+c^2-2(ab+bc+ca)\,,\nonumber\\
A_1&=&2\sin^2\theta_l\big(H_{\frac{1}{2},0}^2+H_{-\frac{1}{2},0}^2\big)+(1-\cos\theta_l)^2H_{\frac{1}{2},+}^2
\nonumber\\&&+(1+\cos\theta_l)^2H_{-\frac{1}{2},-}^2\,,\nonumber\\
A_2&=&2\cos^2\theta_l\big(H_{\frac{1}{2},0}^2+H_{-\frac{1}{2},0}^2\big)+\sin^2\theta_l \big{(}H_{\frac{1}{2},+}^2\nonumber\\&&  +H_{-\frac{1}{2},-}^2\big{)}+2\big(H_{\frac{1}{2},t}^2+H_{-\frac{1}{2},t}^2\big)\nonumber\\&&
-4\cos\theta_l \big(H_{\frac{1}{2},0}H_{\frac{1}{2},t}+H_{-\frac{1}{2},0} H_{-\frac{1}{2},t}\big)\,,\nonumber\\
A_3&=&H^{SP^2}_{\frac{1}{2},0}+H^{SP^2}_{-\frac{1}{2},0}\,, \nonumber\\
A_4&=&-\cos\theta_l\big(H_{\frac{1}{2},0}H^{SP}_{\frac{1}{2},0}+H_{-\frac{1}{2},0}H^{SP}_{-\frac{1}{2},0}\big)\nonumber\\&&
+\big(H_{\frac{1}{2},t}H^{SP}_{\frac{1}{2},0}+H_{-\frac{1}{2},t}H^{SP}_{-\frac{1}{2},0}\big)\,,\nonumber
\end{eqnarray}
}
the  $\theta_l$ is the angle between the directions of the parent baryon $B_1$ and final lepton $l$ three momentum vector
in the dilepton rest frame.

After integrating over the ${\rm cos}~\theta_l$ of Eq.~(\ref{Eq.dTdsdscos}), we can obtain  the normalized differential decay rate
\begin{eqnarray}\label{Eq.dTds}
\frac{d\Gamma(B_1\to  B_2 l \bar{\nu}_l)}{dq^2}&=&\frac{8N}{3}\left(1-\frac{m^2_l}{q^2}\right)^2[\mathcal{B}_1+\frac{m^2_l}{2q^2}\mathcal{B}_2\nonumber\\&&+\frac{3}{2}\mathcal{B}_3+\frac{3m_l}{\sqrt{q^2}}\mathcal{B}_4],
\end{eqnarray}
with
\begin{eqnarray}
\mathcal{B}_1 & =& H^2_{\frac{1}{2}0}+H^2_{-\frac{1}{2}0} +H^2_{\frac{1}{2}+}+H^2_{-\frac{1}{2}-},\nonumber\\
\mathcal{B}_2 & =& H^2_{\frac{1}{2}0}+H^2_{-\frac{1}{2}0} +H^2_{\frac{1}{2}+}+H^2_{-\frac{1}{2}-}+3\left(H^2_{\frac{1}{2}t}+H^2_{-\frac{1}{2}t}\right),\nonumber\\
\mathcal{B}_3 & =& \left(H^{SP}_{\frac{1}{2}0}\right)^2 +\left(H^{SP}_{-\frac{1}{2}0}\right)^2,\nonumber\\
\mathcal{B}_4 &=& H_{\frac{1}{2}t}H^{SP}_{\frac{1}{2}0}+H_{-\frac{1}{2}t}H^{SP}_{-\frac{1}{2}0}.\nonumber
\end{eqnarray}

Besides the differential decay rate,  other interesting observables  are also investigated and they can be written as follows:
\begin{itemize}
	\item[*]
	The total differential branching fraction
	\begin{eqnarray}\label{Eq.dBrds}
	\frac{d\mathcal{B}(B_1\to  B_2 l \bar{\nu}_l)}{dq^2}&=&\tau_{\Omega_b(\Sigma_b)}\frac{d\Gamma(B_1\to  B_2 l \bar{\nu}_l)}{dq^2}.
	\end{eqnarray}
	\item[*] The lepton side forward-backward asymmetries parameter
	\begin{eqnarray}\label{Eq.AFBds}
	A_{\rm FB}^l(q^2)&=&\bigg{ (}  \int_{-1}^0 d \cos \theta_l \frac{d^2 \Gamma}{d q^2 d \cos \theta_l}\nonumber\\&& - \int_0^1 d \cos \theta_l \frac{d^2 \Gamma}{d q^2 d \cos \theta_l} \bigg{)} \Big {/}\frac{d \Gamma}{d q^2}\,.
	\end{eqnarray}
	\item[*] The convexity parameter
	\begin{eqnarray} \label{Eq.CFlds}
	C_F^l(q^2) = \frac{1}{ d \Gamma/d q^2} \frac{\,d^2}{d (\cos\theta_l) ^2\,}\Bigg(\frac{d^2 \Gamma}{dq^2 d\cos\theta_l}\Bigg).
	\end{eqnarray}
	\item[*] The longitudinal  polarization asymmetries parameter of daughter baryons $\Omega_c(\Sigma_c)$
	\begin{eqnarray}\label{Eq.PLhds}
	P_L^{\Omega_c(\Sigma_c)}(q^2)=\frac{{\rm d}\Gamma^{\lambda_2=1/2}/{\rm d}q^2-
		{\rm d}\Gamma^{\lambda_2=-1/2}/{\rm d}q^2}{{\rm d}\Gamma/{\rm d}q^2}\,,
	\end{eqnarray}
	where ${\rm d}\Gamma^{\lambda_2=\pm 1/2}/dq^2$ are the individual helicity dependent differential decay rates,
whose detailed expressions are given in Ref.~\cite{Ray:2018hrx}.
	\item[*] The longitudinal polarization asymmetries parameter of the charged lepton
	\begin{eqnarray}\label{Eq.PLlds}
	P_L^{l}(q^2)=\frac{{\rm d}\Gamma^{\lambda_{l}=1/2}/{\rm d}q^2-
		{\rm d}\Gamma^{\lambda_{l}=-1/2}/{\rm d}q^2}{{\rm d}\Gamma/{\rm d}q^2}\,,
	\label{Pol_lept}
	\end{eqnarray}
	where ${\rm d}\Gamma^{\lambda_l=\pm 1/2}/dq^2$  are differential decay rates for positive and negative helicity of lepton and
their  detailed expressions  are also  given in Ref.~\cite{Ray:2018hrx}.
	\item[*] The ratios of the branching fractions
	\begin{eqnarray}\label{Eq.RB2}
	{\rm R}_{\Omega_c(\Sigma_c)}(q^2)=\frac{d\mathcal{B}(B_1\to B_2 \tau\bar{\nu}_{\tau})/dq^2}{d\mathcal{B}(B_1\to B_2 \ell\bar{\nu}_{\ell})/dq^2}.
	\end{eqnarray}
\end{itemize}
Note that integrating the numerator and denominator over $q^2$ separately before taking the ratio,
we can get the average values of all the observables such as
$\langle A_{\rm FB}^l\rangle$,  $\langle C_{F}^l\rangle$,  $\langle P_{L}^l\rangle$, $\langle P_{L}^{\Omega_c(\Sigma_c)}\rangle$ and  $\langle {\rm R}_{\Omega_c(\Sigma_c)}\rangle$.

\section{Numerical analysis and discussion}\label{Sec.result}
In this section, we will give our results within SM and various NP scenarios in a model independent way.
We present the constrained  NP coupling parameter space and give the numerical results of the observables displayed in
Eqs.~(\ref{Eq.dBrds})-(\ref{Eq.RB2}) for $\Omega_b\to \Omega_c l \bar{\nu}_l$ and $\Sigma_b\to \Sigma_c l \bar{\nu}_l$ transitions including the contributions of different NP coupling parameters.
In order to get the allowed NP coupling parameter  space in various NP scenarios, we will impose the $2\sigma$ constraint coming from the latest experimental values of the observables $\mathcal{B}(B\to D^{(*)}l \bar{\nu}_l)$, ${\rm R}_{D^{(*)}}$ and ${\rm R}_{J/\psi}$.
The specific expressions of these observables for $B\to D^{(*)} l \bar{\nu}_l$ and $B_c\to J/\psi l\bar{\nu}_l$  processes used in our work can easily be found in  the  Refs.~\cite{Ray:2018hrx,Zhang:2019xdm,Dutta:2013qaa,Sakaki:2013bfa,Dutta:2016eml}.


In our numerical computation about above various observables, except for the transition form factors and the NP coupling parameters, the values of the other input parameters such as the particle masses, decay constants,
mean lives and some relevant experimental measurement data  of $\mathcal{B}(B\to D^{(*)}l \bar{\nu}_l)$ are mainly taken from the Particle Data Group (PDG)~\cite{PDG2020}.
The relevant experimental data about ${\rm R}_{D^{(*)}}$ and ${\rm R}_{J/\psi}$ used in this work are listed in Eqs.~(\ref{Eq.EXRD}) and (\ref{Eq.EXRJP}).
Note that, in the model independent analysis, we assume that all the NP coupling parameters are complex and  we consider only one  NP coupling existing in
Eq.~(\ref{Eq.Lbtoclnul}) at one time and keep it interference with the SM.

\begin{figure*}[ht]
	\centering
	\includegraphics[width=4cm,height=4.0cm]{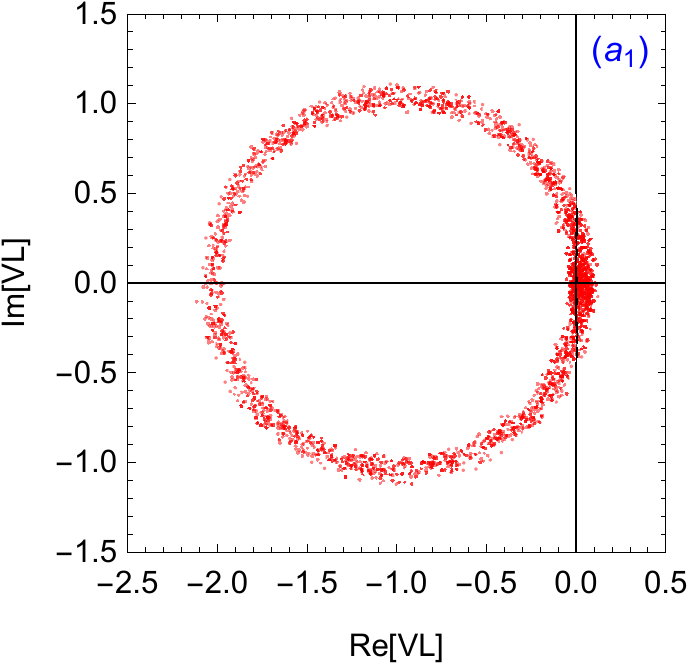}
	\includegraphics[width=4cm,height=4.0cm]{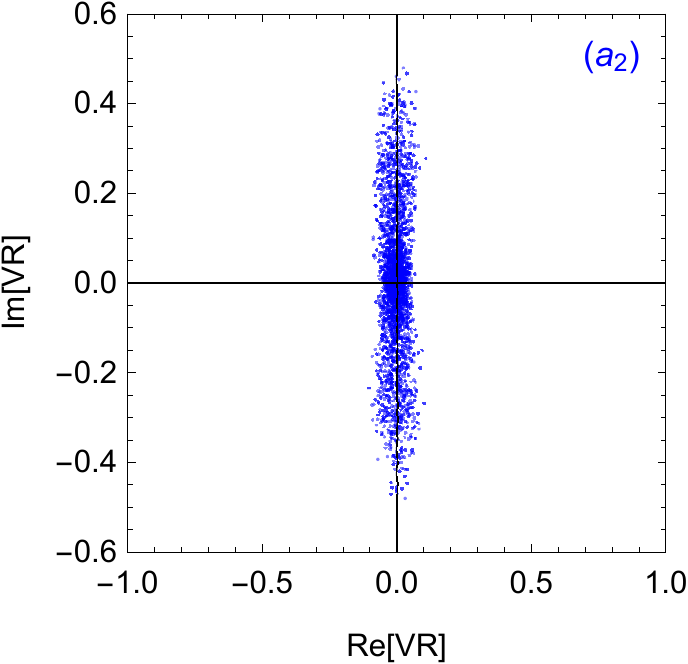}
	\includegraphics[width=4cm,height=4.0cm]{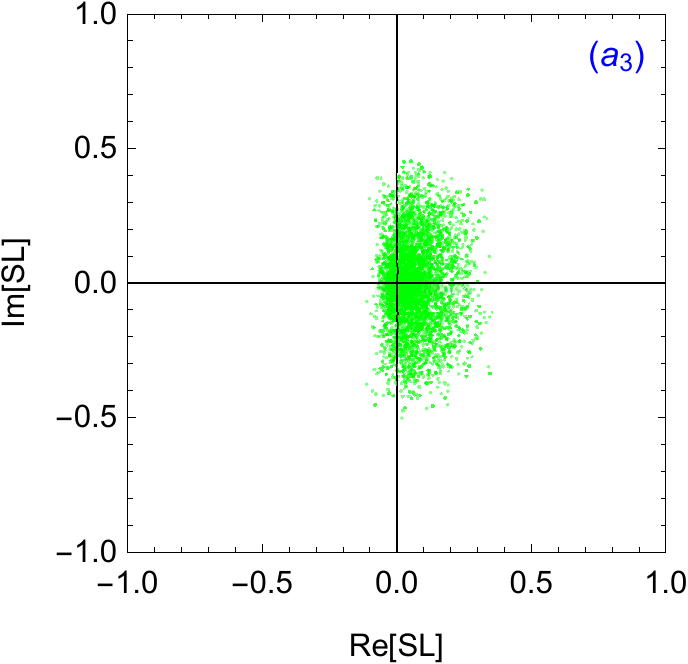}
	\includegraphics[width=4cm,height=4.0cm]{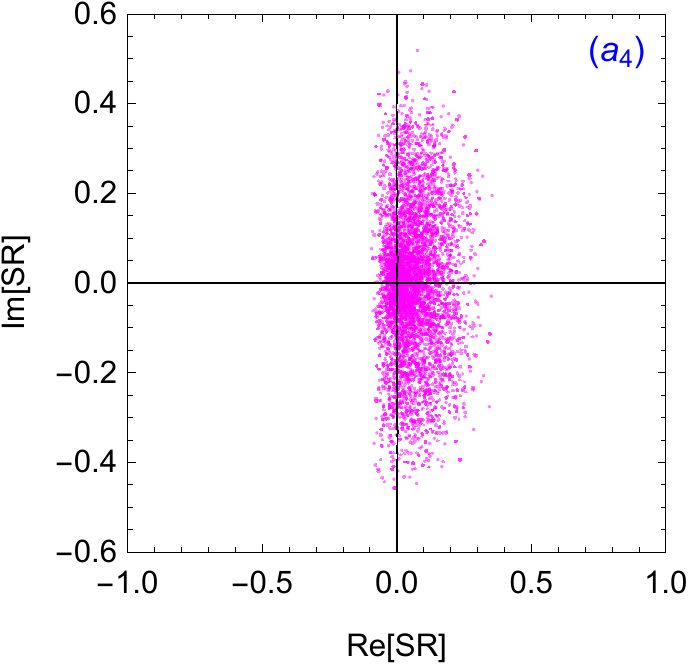}\\
   ~\includegraphics[width=4cm,height=4.0cm]{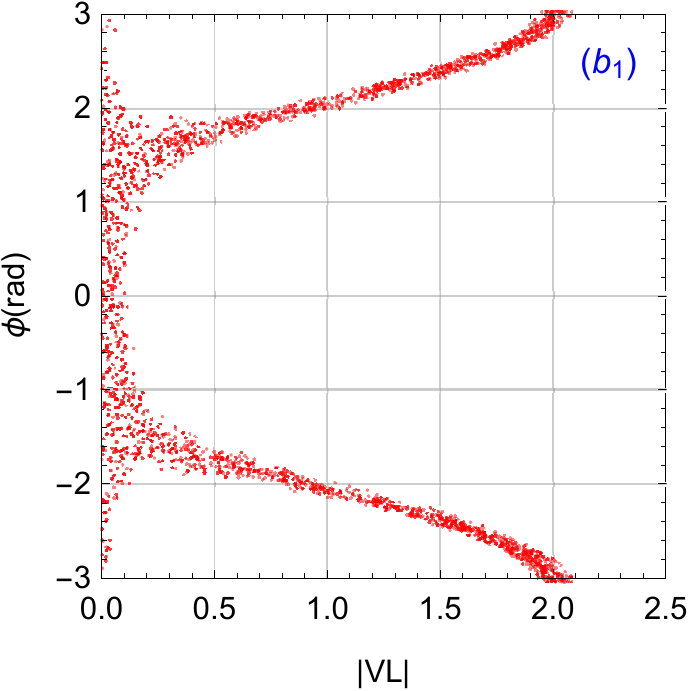}
	\includegraphics[width=4cm,height=4.0cm]{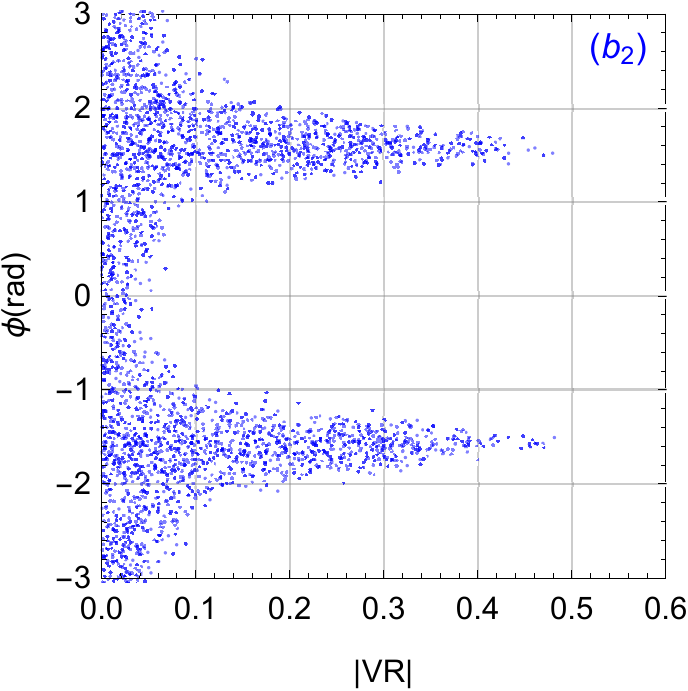}
	\includegraphics[width=4cm,height=4.0cm]{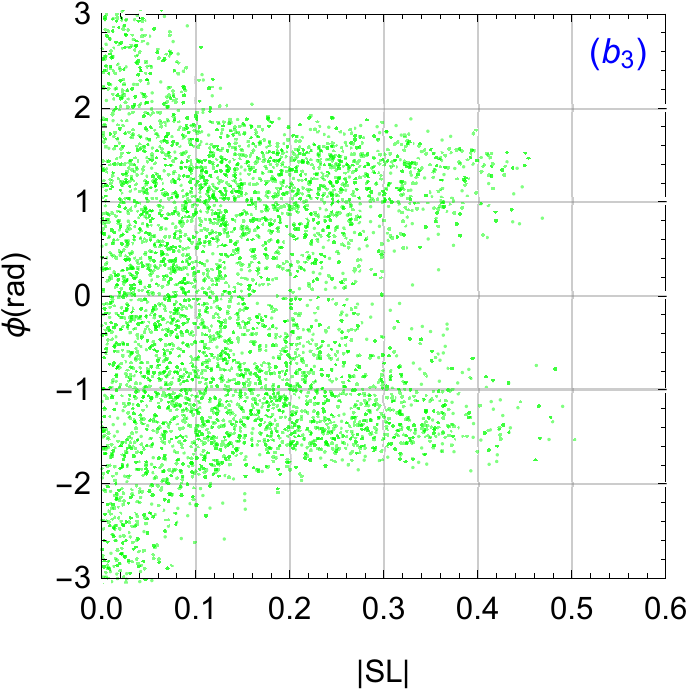}
	\includegraphics[width=4cm,height=4.0cm]{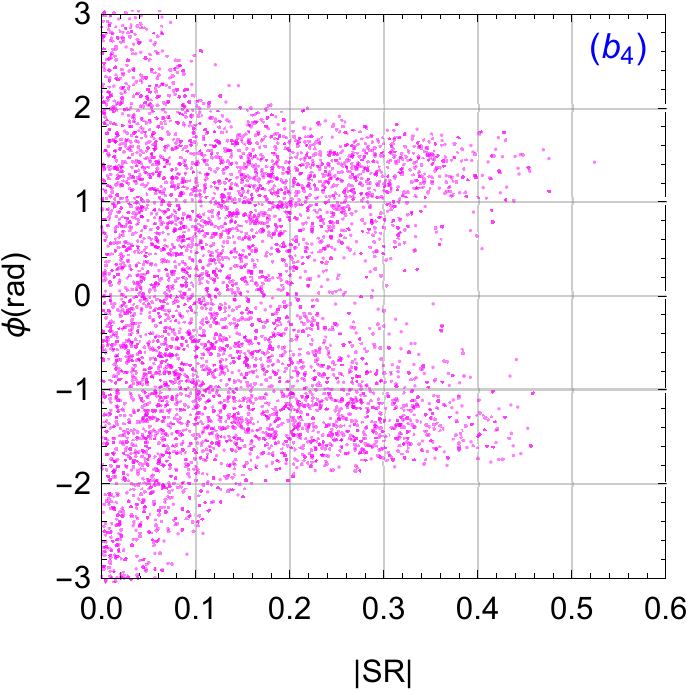}\\
	\caption{  {\small The bounds on both real-imaginary ($a_1$-$a_4$)  and modulus-phase ($b_1$-$b_4$)
parts of the complex  coupling parameters VL,VR, SL and SR
coming from the relevant experimental constraints.}}
	\label{fig.limitNP}
\end{figure*}
\begin{figure*}[ht]
	\centering
	\includegraphics[width=4cm,height=4.0cm]{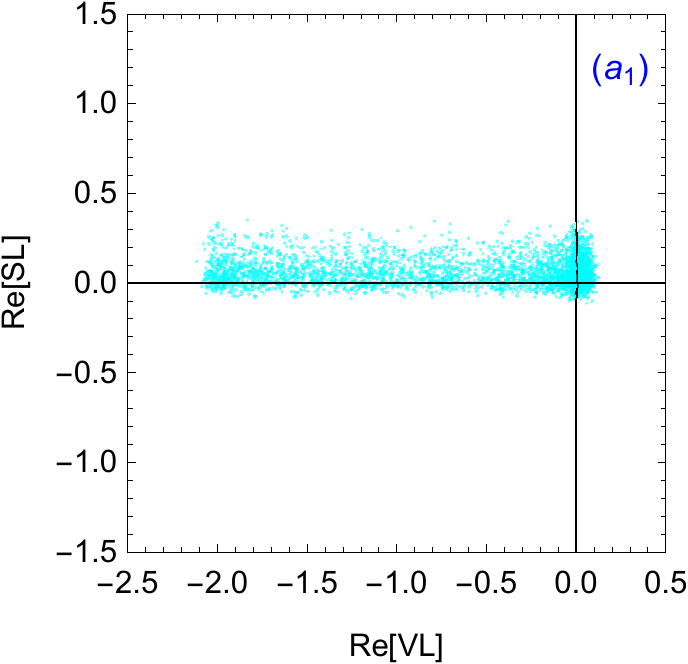}
	\includegraphics[width=4cm,height=4.0cm]{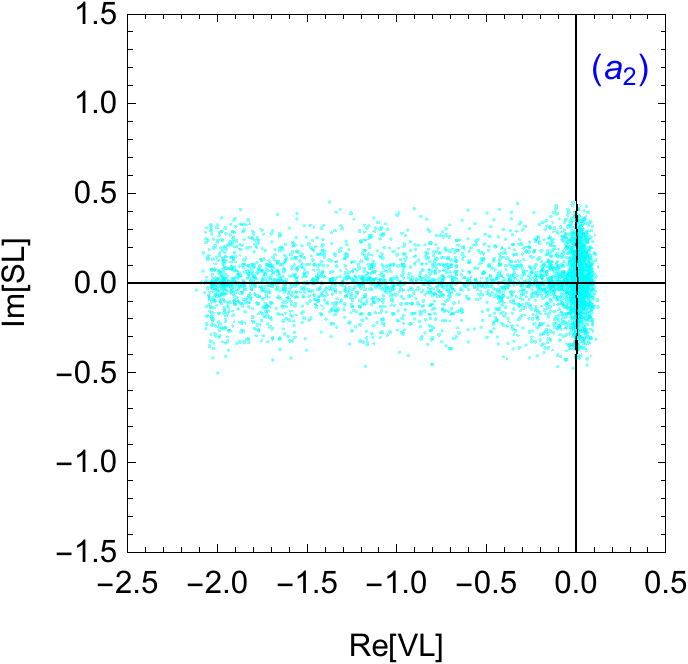}
	\includegraphics[width=4cm,height=4.0cm]{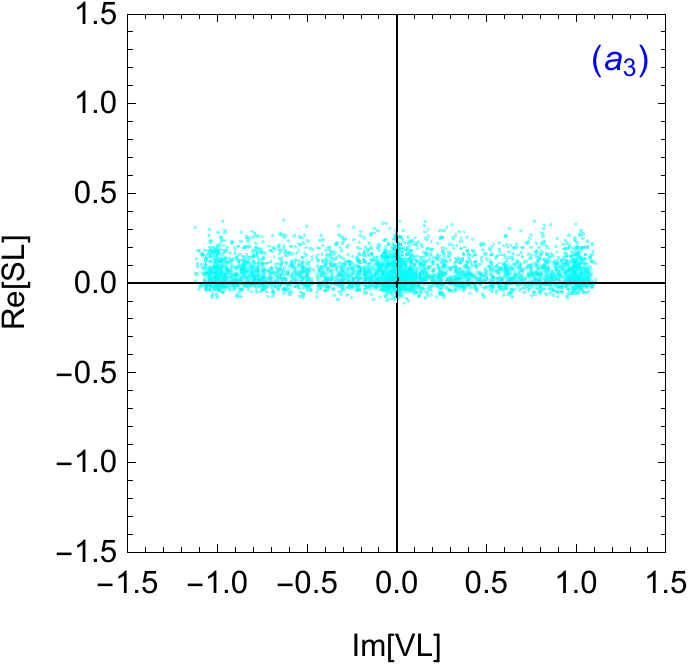}
	\includegraphics[width=4cm,height=4.0cm]{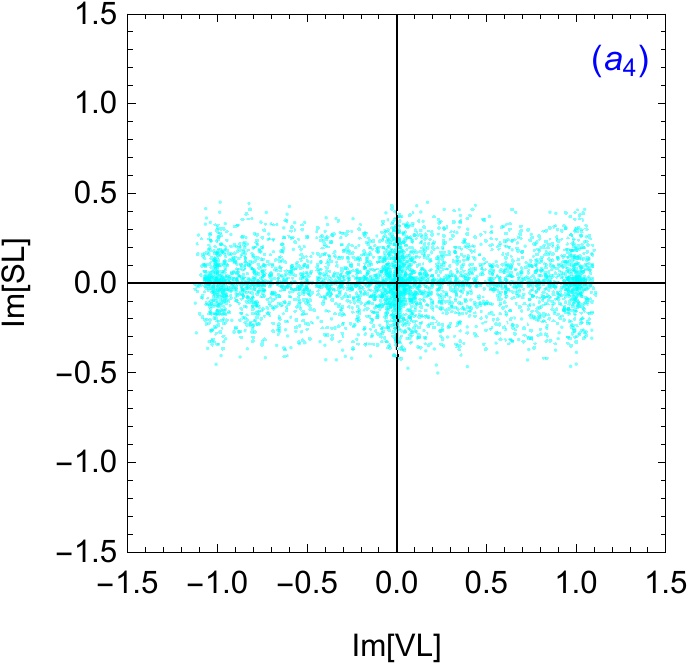}\\
	\includegraphics[width=4cm,height=4.0cm]{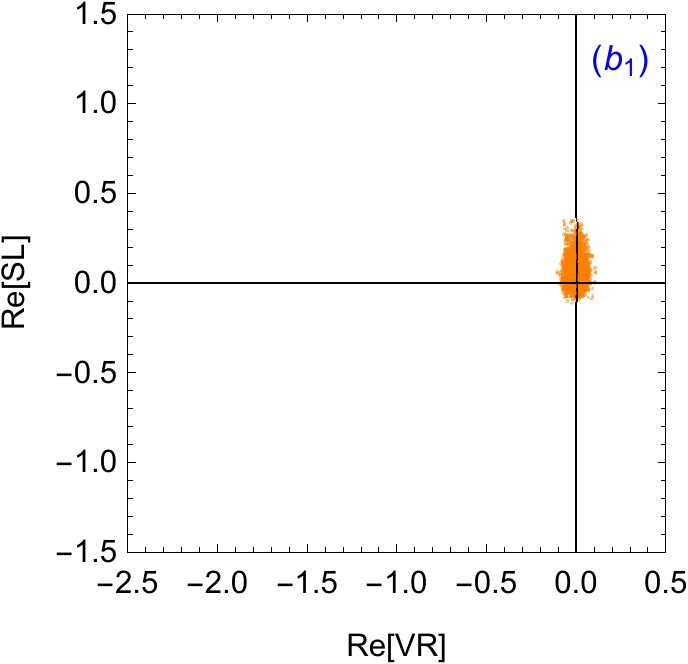}
	\includegraphics[width=4cm,height=4.0cm]{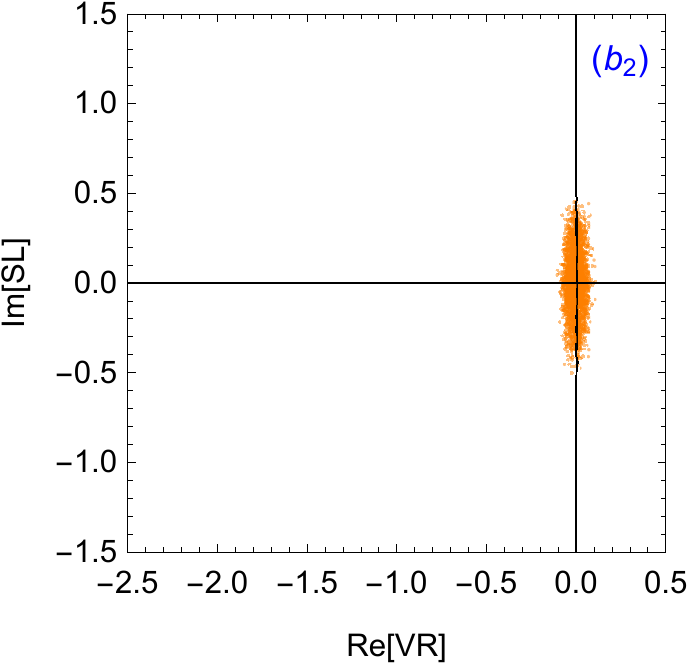}
	\includegraphics[width=4cm,height=4.0cm]{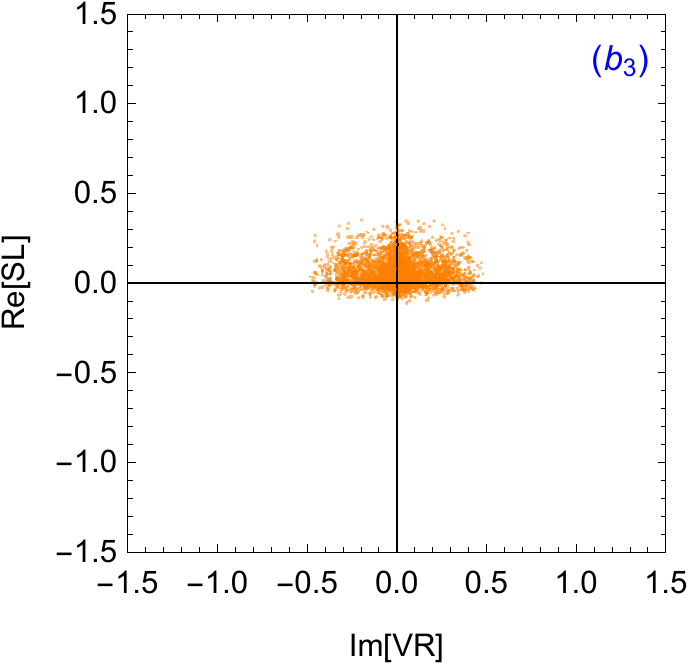}
	\includegraphics[width=4cm,height=4.0cm]{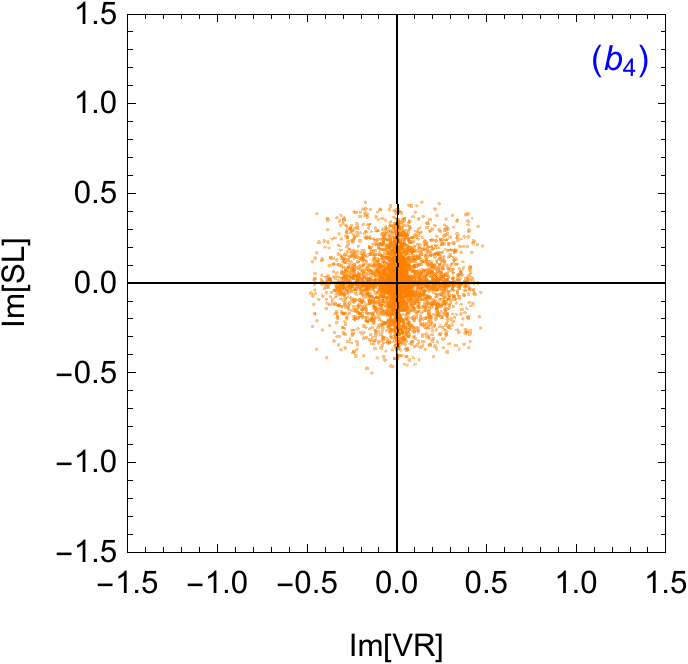}\\
	\caption{  {\small The bounds on both real and imaginary parts of the complex coupling parameters VL and  SL ($a_1-a_4$),
VR and  SL ($b_1-b_4$) coming from the relevant experimental results.}}
	\label{fig.NPcouplingReIm}
\end{figure*}

Firstly, we obtain the constrained range of NP coupling parameters $V_L$, $V_R$, $S_L$ and $S_R$ by using the recent experimental measurement results,
and then examine the  NP  effects on the observables which are displayed  in Sec.~\ref{Sec.two} by using the constrained NP coupling parameters.
The constrained the range of four NP coupling parameters $V_L$, $V_R$, $S_L$ and  $S_R$ are shown in the Fig.~\ref{fig.limitNP},
and the results can be intuitively displayed by both real-imaginary and modulus-phases of the NP coupling parameters in the figure.
There are few references that discuss the relationship between modulus and phases of the NP coupling parameters.
The constrained  results on the real, imaginary and modulus of the NP coupling parameters are  listed in the Tab.~\ref{Tab:NPlimit} clearly.
From Fig.~\ref{fig.limitNP} we can see that present experimental data give quite strong bounds on the relevant coupling parameters, in particular, modulus and phase
of $V_L$ is strongly restricted.
The constrained range of $V_L$ and $S_L$, $V_R$ and $S_L$ are shown
in Fig.~\ref{fig.NPcouplingReIm} (a1-a4) and (b1-b4), respectively.
From Fig.~\ref{fig.NPcouplingReIm} (a1-a4) we can see that the values of ${\rm Re}[S_L]$ and ${\rm Im}[S_L]$ are in small range compared with the values of ${\rm Re}[V_L]$ and ${\rm Im}[V_L]$.
From Fig.~\ref{fig.NPcouplingReIm} (b1-b4),  it is clear to find the result of $V_R$-$S_L$ presents an axial symmetric phenomenon,
and the  scattered points are mainly distributed around the origin.
Because the distribution relationship of $V_L$-$S_R$ and  $V_R$ -$S_R$ are similar to the Fig.~\ref{fig.NPcouplingReIm},
we do not show the relationship of $V_L$-$S_R$ and  $V_R$ -$S_R$  anymore.

{\footnotesize\renewcommand\arraystretch{1.25}
\begin{table*}[htb]
	\centering
	\caption{ {\small The allowed ranges of $V_L$, $V_R$, $S_L$ and $S_R$ NP coupling coefficients}.} \label{Tab:NPlimit}
	\begin{tabular}{c|c|c|c|c}
         \hline
		\hline
		Decay mode~&~NP coefficients~ & ~Min value~ & ~Max Value~  &~ Max of $|V_i(S_i)|(i=L.R)$\\
		\hline
		~&~$({\rm Re}[V_L], {\rm Im}[V_L]) $~&~$(-2.116,-1.123)$~&~$(0.121,1.109)$&2.118\\
		$b\to c l\bar{\nu_{l}} $~&~$({\rm Re}[V_R], {\rm Im}[V_R]) $~&~$(-0.105,-0.481)$~&~$(0.105,0.479)$&0.482\\
		~&~$({\rm Re}[S_L], {\rm Im}[S_L]) $~&~$(-0.111,-0.502)$~&~$(0.351,0.451)$&0.502\\
		~&~$({\rm Re}[S_R], {\rm Im}[S_R]) $~&~$(-0.094,-0.456)$~&~$(0.355,0.519)$&0.524\\
		\hline\hline
	\end{tabular}
\end{table*}
}

The constraints about these NP coupling parameters  obtained from various B meson decay processes have been also discussed in Refs.~\cite{Ray:2019gkv,Ray:2018hrx,Shivashankara:2015cta,Zhang:2019xdm,Ivanov:2016qtw,Ivanov:2017mrj,Dutta:2015ueb}.
The NP coupling parameters are assumed complex or real in these references
 and corresponding experimental data  which are used in these references are mainly from ${\rm R}_{D^{(*)}}$ and ${\rm R}_{J/\psi}$.
But few references consider the experimental data of  $\mathcal{B}(B\to D^{(*)} l \bar{\nu}_{l})$ which are considered in our work.
In our analysis, we use  the experimental data of ${\rm R}_{D^{(*)}}$, ${\rm R}_{J/\psi}$ and $\mathcal{B}(B\to D^{(*)} l \bar{\nu}_{l})$ to constrain the space of the corresponding NP coupling parameters.
We get more severe bounds on the phases and strengths of the NP coupling parameters and we also give the relationship between  modulus and phase of four NP coupling parameters which are not discussed in many previous references.

Employing the theoretical framework described in Sec.~\ref{Sec.two},
the SM predictions  are reported for processes $\Omega_b\to \Omega_c l \bar{\nu}_l$ and $\Sigma_b\to \Sigma_c l \bar{\nu}_l$.
In Tab.~\ref{Tab:smcentral}, we list the average values of $\Gamma$, $\langle P_L^l \rangle$, $\langle P_L^{\Omega_c(\Sigma_c)} \rangle$, $\langle A_{FB}^l \rangle$, $\langle C_{F}^l \rangle$ and $\langle {\rm R}_{\Omega_c(\Sigma_c)}\rangle$ for $e$, $\mu$ and $\tau$ mode respectively.
From Tab.~\ref{Tab:smcentral}, one can see that the results for $e$ mode and $\mu$ mode are close for $\Omega_b\to \Omega_c l \bar{\nu}_l$ and $\Sigma_b\to \Sigma_c l \bar{\nu}_l$ processes.
The total decay rates $\Gamma$ (in units of $10^{10} s^{-1}$ )  at $l=e,\mu$ are observed to be larger than the result at $l=\tau$,
and  same phenomenon arises in $\langle P_L^{\Omega_c(\Sigma_c)} \rangle$ and $A_{FB}^l$.
The lepton polarization fractions $P_L^l$ for the $e$ and $\mu$ are negative, but one for the $\tau$ mode is positive.
The forward-backward asymmetries $A_{FB}^l$ for  $e$ and $\mu$  mode are positive, but one of the $\tau$ mode is negative.
The hadron polarization fractions $P_L^{\Omega_c(\Sigma_c)}$  are about 0.58 at $l=e,\mu$,
and the result is about 0.35 at $l=\tau$ for both $\Omega_b\to \Omega_c l \bar{\nu}_l$ and $\Sigma_b\to \Sigma_c l \bar{\nu}_l$.
All the convexity parameters $\langle C_{F}^l \rangle$  are negative and $\langle C_{F}^{\tau} \rangle$ is much larger than $\langle C_{F}^{l} \rangle$ ($l=e,\mu$).
The ratio of branching ratio $\langle {\rm R}_{\Omega_c}\rangle$ is slightly larger than
$\langle {\rm R}_{\Sigma_c}\rangle$.

\begin{figure*}[ht]
	\centering
	\includegraphics[width=5.5cm,height=5cm]{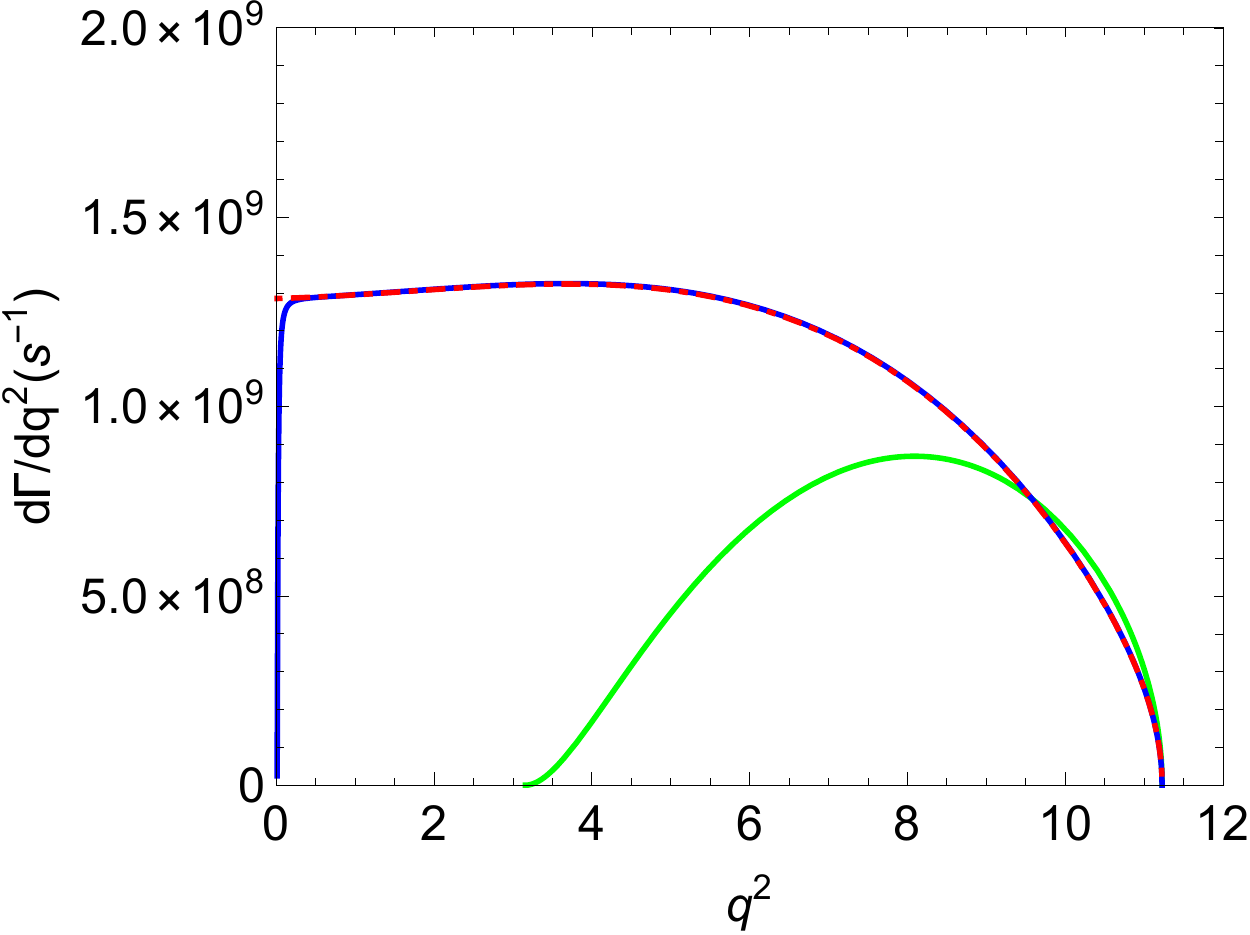}~
	\includegraphics[width=5.5cm,height=5.0cm]{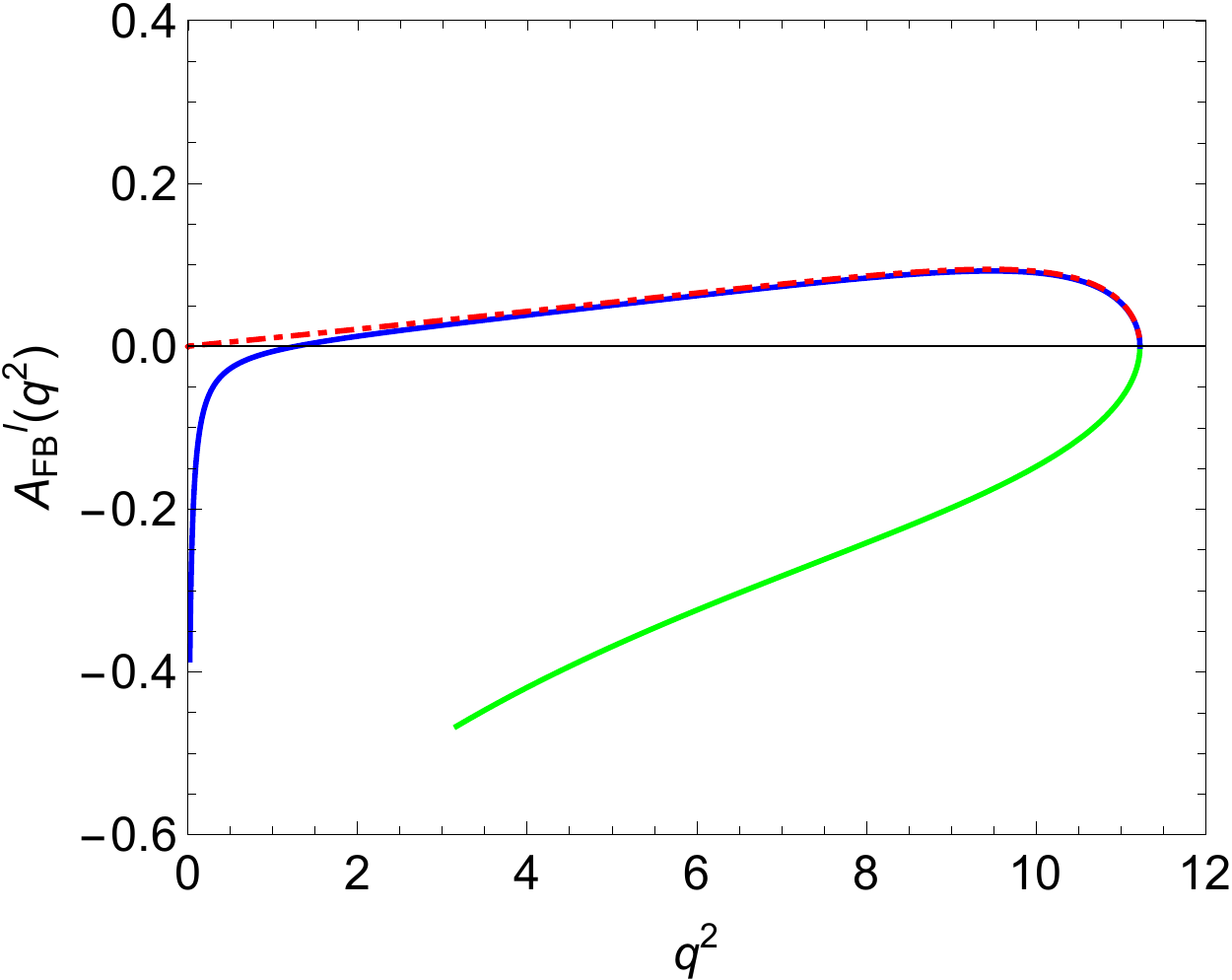}
	\includegraphics[width=5.5cm,height=5.0cm]{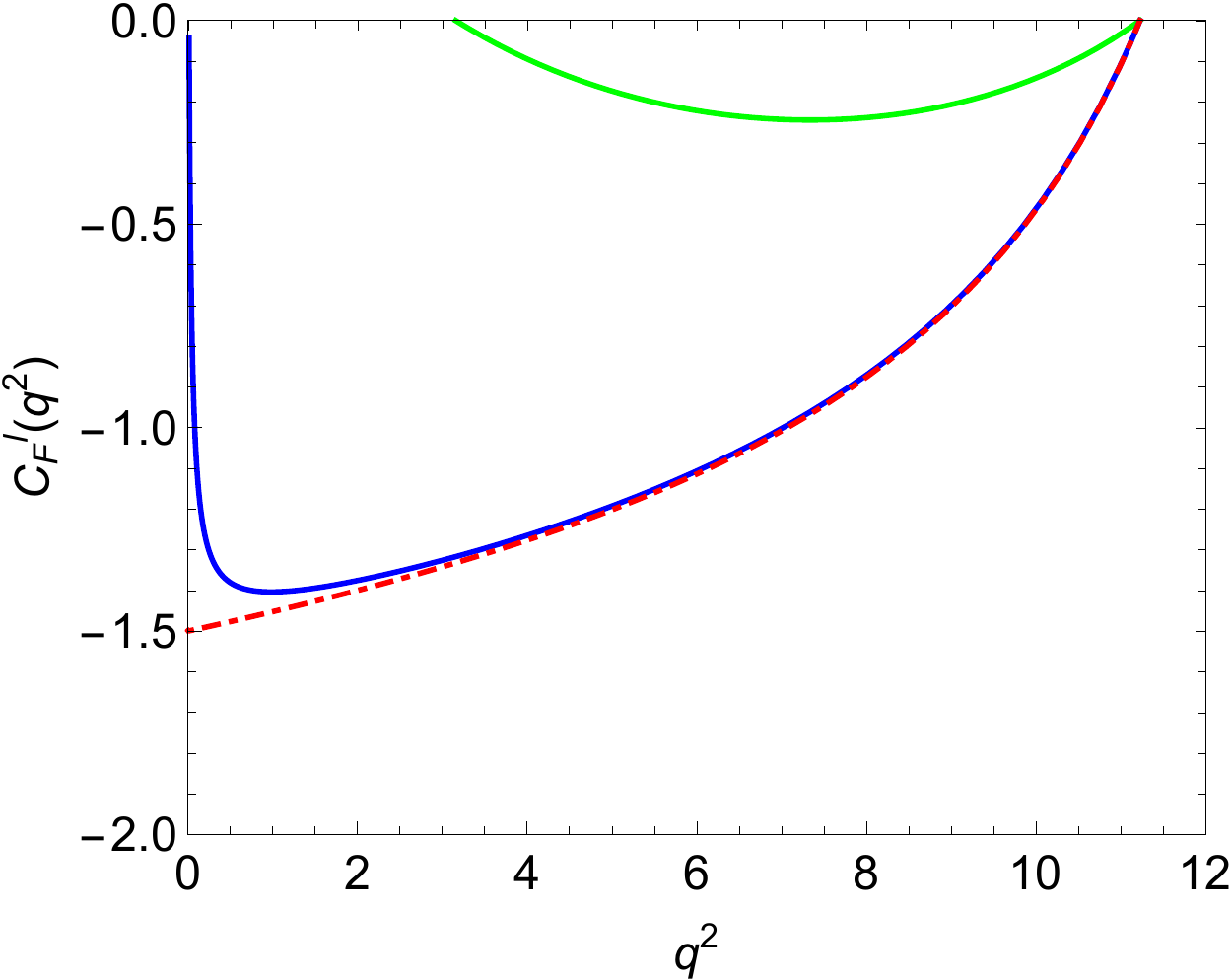}\\
	\includegraphics[width=5.5cm,height=5.0cm]{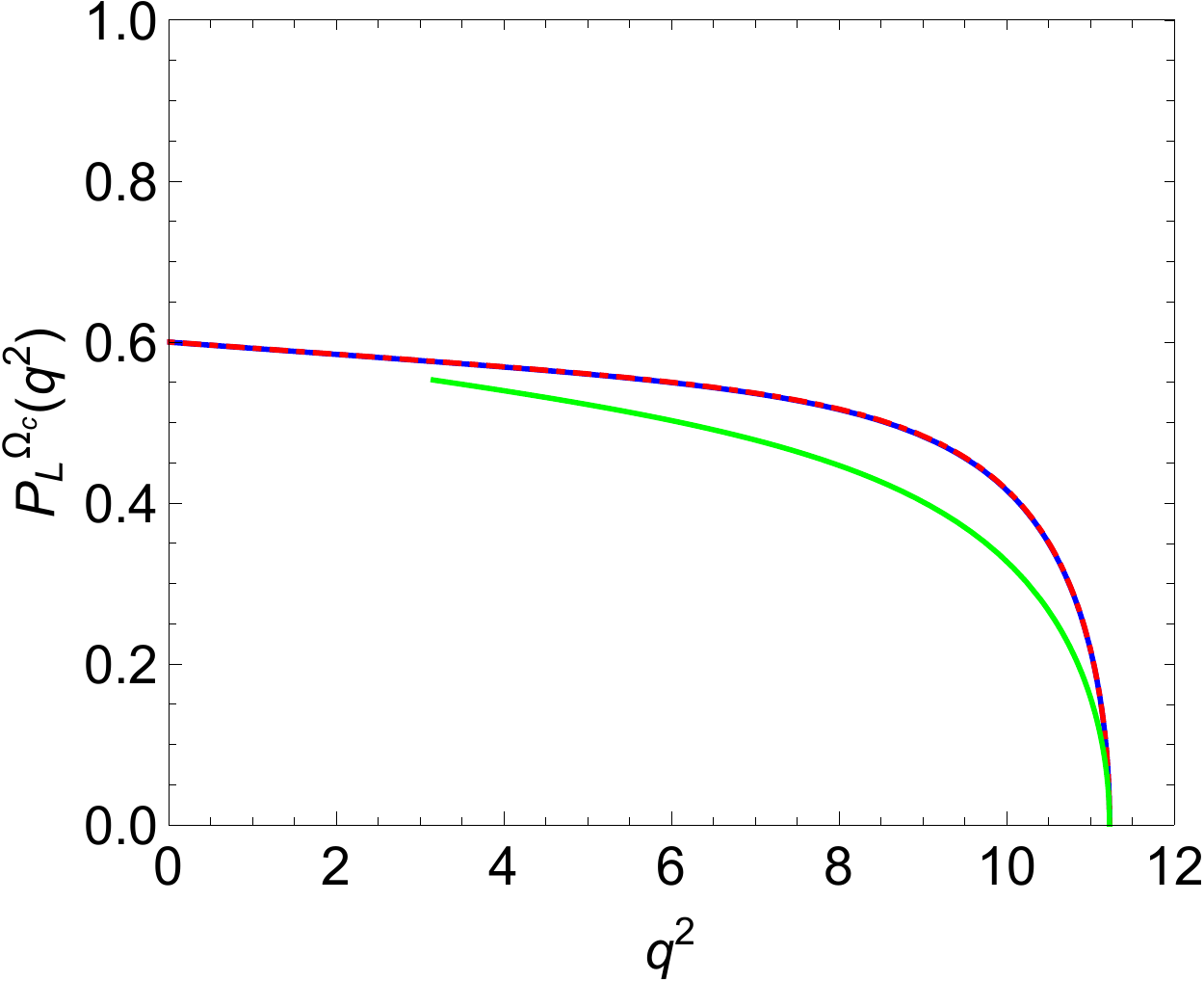}
	\includegraphics[width=5.5cm,height=5.0cm]{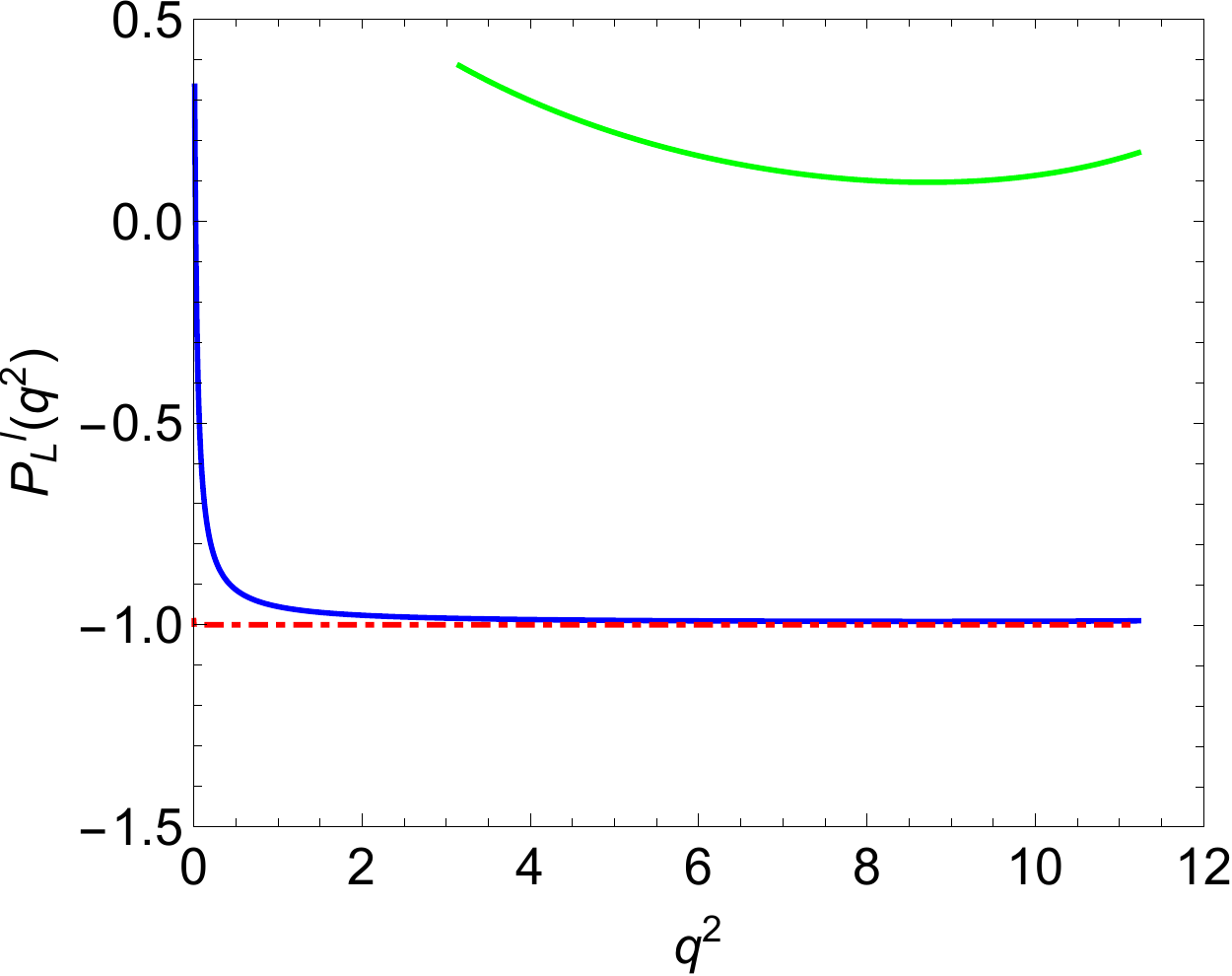}
	\includegraphics[width=5.5cm,height=5.0cm]{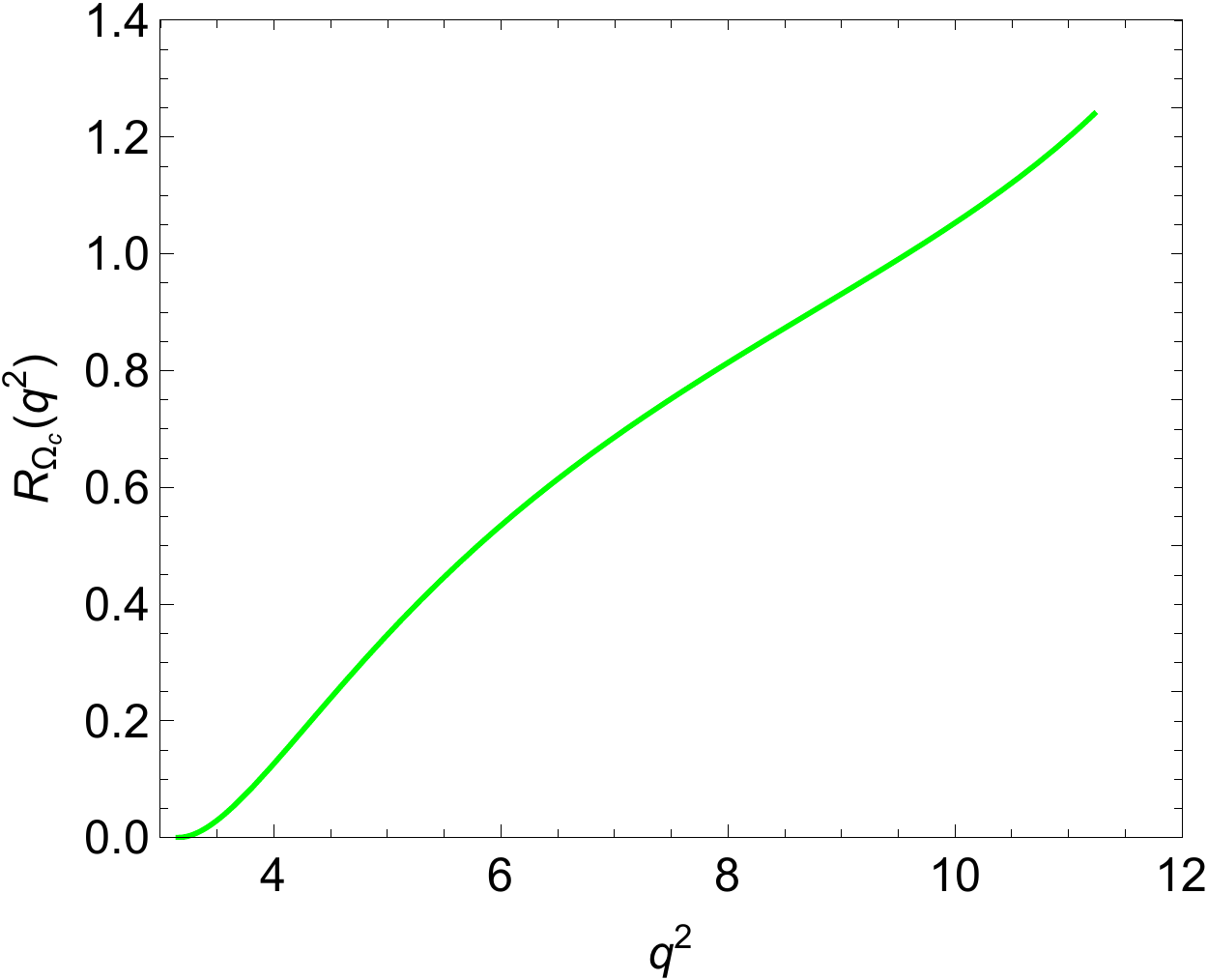}\\
	\caption{  {\small The SM predictions for the $q^2$ dependent observables $d\Gamma/dq^2$, $A_{FB}^l(q^2) $, $C_F^l(q^2)$,
$P_L^{\Omega_c}(q^2)$, $P_L^l(q^2)$ and ${\rm R}_{\Omega_c}(q^2)$ relative  to the decays
 $\Omega_b\to \Omega_c l \bar{\nu_l}$ $(\ell=e,\mu,\tau)$. The red dot dash line, blue and green line
 represent the $e$, $\mu$ and  $\tau$ mode,  respectively.  }}\label{fig.ObSM}
\end{figure*}

The behaviors of each observable as a function of $q^2$ for the processes $\Omega_b\to \Omega_c l \bar{\nu_l}$ and $\Sigma_b\to \Sigma_c l \bar{\nu}_l$  are similar to each other.
So we only take  $\Omega_b\to \Omega_c l \bar{\nu_l}$ decays as an example to illustrate in detail and the same goes in the following text.
 The SM predictions for the $q^2$ dependency of different observables  in the reasonable kinematic range for $\Omega_b\to \Omega_c l \bar{\nu_l}$  are displayed in Fig.~\ref{fig.ObSM}.
In this figure, we compare the distributions of the each observable and the red dot dash line, blue and green line represents the $e$, $\mu$ and $\tau$ mode, respectively.
The $q^2$ dependency of $d\Gamma/dq^2$, $A_{FB}^{l}$, $C_F^{l}$ and $P_L^{l}$ are distinct for three generation leptons.
But we can find that the variation tendency
 of  $d\Gamma/dq^2$, $A_{FB}^{l}$, $C_F^{l}$ and $P_L^{l}$ for $e$ and $\mu$ modes is  almost same except in small $q^2$ region.
The total differential decay rate for $e$ is maximum at  $q^2_{min}$ and minimum at  $q^2_{max}$,  however, the result for $\tau$ is maximum when $q^2\approx 8 GeV^2$ and approaches zero at  $q^2_{min}$ and $q^2_{max}$.
For $\mu$ mode, $d\Gamma/dq^2$ changes to zero quickly when $q^2=m_{\mu}^2$ due to the effect of $\mu$ mass.
All the  $A_{FB}^{l}$  approach to zero at $q^2_{max}$.
The $A_{FB}^e$ is positive  while $A_{FB}^{\tau}$ is negative and  great increasing with $q^2$ over the all $q^2$ region.
Besides, $A_{FB}^{\mu}$ changes to -0.4 quickly when $q^2=m_{\mu}^2$ and there is a zero-crossing point, which lies in the low $q^2$ region.
All the $C_F^l$ are negative in the whole $q^2$ region and   at the large $q^2$ limit $C_F^{l}$ are zero.
At the low $q^2$ range
$C_F^{e}$ is  around -1.5 when $q^2=q^2_{min}$,
and $C_F^{\mu}\approx-1.4$ when $q^2\approx0.4GeV^2$, while $C_F^{\mu}$ changes to zero quickly when $q^2=m_{\mu}^2$ due to the effect of the lepton mass.
This behavior indicates that the $cos\theta$ distribution in $q^2 \in [0.4,11.23]$ is strongly parabolic.
On the contrary, the  $C_F^{\tau}$ is small in the whole ranges, which implies a straight-line behavior of the $cos\theta$ distribution.
The $P_L^{\Omega_c}$ are zero for three modes at $q^2_{max}$.
The results of $P_L^{\Omega_c}$ for $e$ and $\mu$ modes completely coincide and it is around 0.6 at $q^2=q^2_{min}=m_l^2$.
The $P_L^e$ is -1 over the all $q^2$ region and it is  similar to $\mu $ mode except for low $q^2$ region.
When $q^2=m_{\mu}^2$, the $P_L^{\mu}$ changes to 0.4 quickly.
While for the $\tau$ mode, the behavior is quite different and   $P_L^{\tau}$ take only positive values for entire $q^2$ values.
The ${\rm R}_{\Omega_c}$ show an almost positive slope over the whole $q^2$ region and ${\rm R}_{\Omega_c}$ is around 0 when $q^2=q^2_{min}$.
Because the ${\rm R}_{\Omega_c}$ is  ratios of the differential branching fraction with the heavier $\tau$
in the final state to the differential branching fraction with the lighter lepton  in the final state,
the result of this observable do not distinguish for the different leptons in the final state.

{\small

\begin{table*}[htbp]
	\centering
	\setlength{\tabcolsep}{8pt} 
	\renewcommand{\arraystretch}{1.5} 
	\caption{	{\small The SM central values for the decay rate $\Gamma$, the lepton polarization fraction $\langle P_L^l \rangle$,
the hadron  polarization fraction 	$\langle P_L^{\Sigma_c(\Omega_c)} \rangle$, the forward-backward asymmetry $\langle A_{FB}^l \rangle$,
the convexity factor $\langle C_{F}^{l} \rangle$ and  the ratio of branching ratio $\langle {\rm R}_{\Sigma_c(\Omega_c)} \rangle$
			for the $e$ mode, $\mu$ mode and $\tau$ mode of $\Omega_b\to\Omega_c l \bar{\nu_l}$ and $\Sigma_b\to\Sigma_c l \bar{\nu_l}$  decays.}}
	\begin{tabular}{cccc|ccc}
		\hline
		\hline
		& &\multicolumn{2}{c}{$\Omega_b \to \Omega_c l \nu$} &  &{$\Sigma_b \to \Sigma_c l \nu$}\\
		\cline{2-7}
		&	$e$ mode&$\mu$ mode&$\tau$ mode&~~~~$e$ mode&$\mu$ mode&$\tau$ mode\\
		\hline
		$ \Gamma \times 10^{10}$ s$^{-1}$& 1.295 & 1.292&0.529 & 1.610 & 1.641 &0.540\\
		\hline
		$\langle P_L^l \rangle$ &  -1.123 & -1.093 &0.135 &-1.135  & -1.131 &0.132\\
		\hline
		$\langle P_L^{\Omega_c(\Sigma_c)} \rangle$ &  0.586 & 0.585 &0.354 & 0.582 & 0.582 &0.355\\
		\hline
		$\langle A_{FB}^l \rangle$& 0.062 & 0.052&-0.220 & 0.065 & 0.055 & -0.220\\
		\hline
		$\langle C_{F}^{l} \rangle$& -1.170 &-1.140 &-0.135 &-1.178  &-1.148  &-0.139 \\
		\hline
		$\langle {\rm R}_{\Omega_c(\Sigma_c)} \rangle$   &\multicolumn{2}{c}{${\rm R}_{\Omega_c}=0.370$}& && ${\rm R}_{\Sigma_c}=0.339$ \\
		\hline
		\hline
	\end{tabular}\label{Tab:smcentral}
\end{table*}

}

\begin{figure*}[ht]
	\centering
	\includegraphics[width=3.9cm,height=3.0cm]{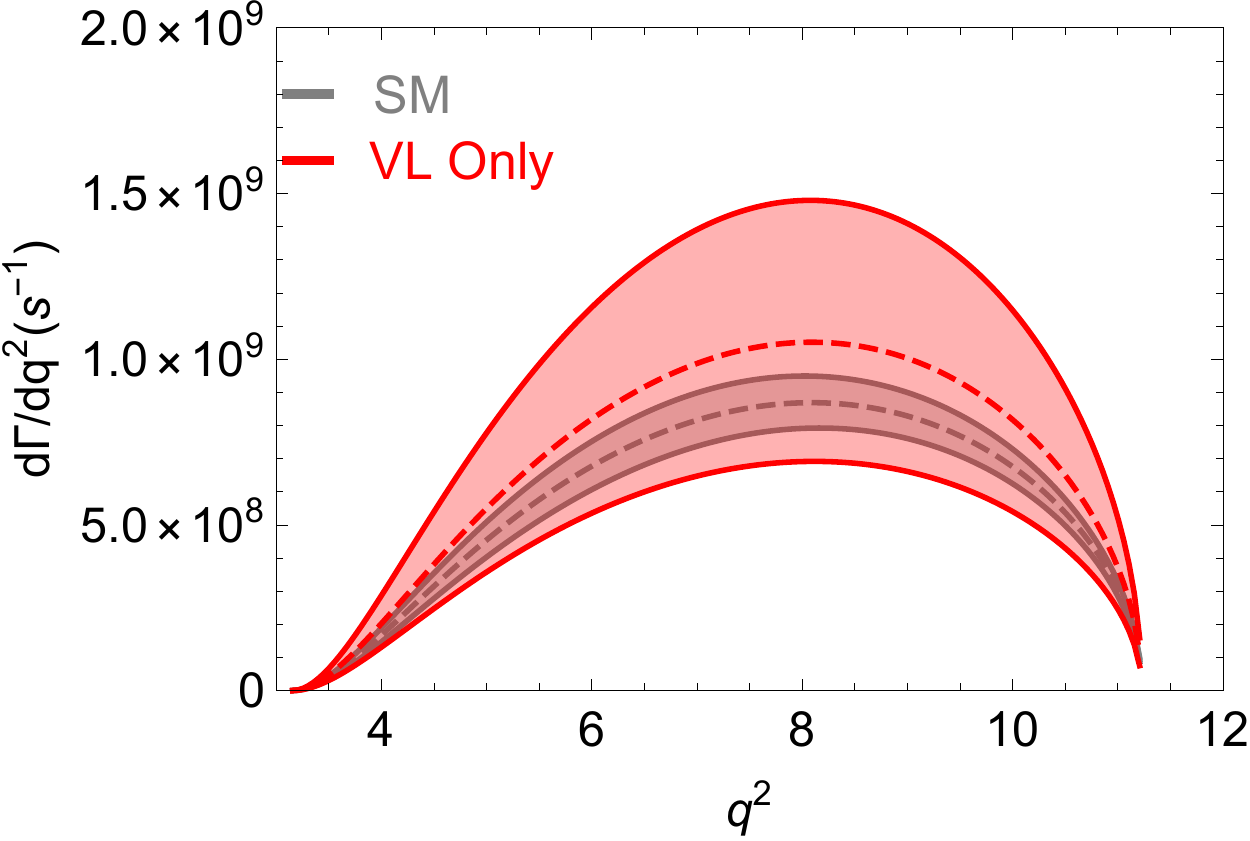}~
	\includegraphics[width=3.9cm,height=3.0cm]{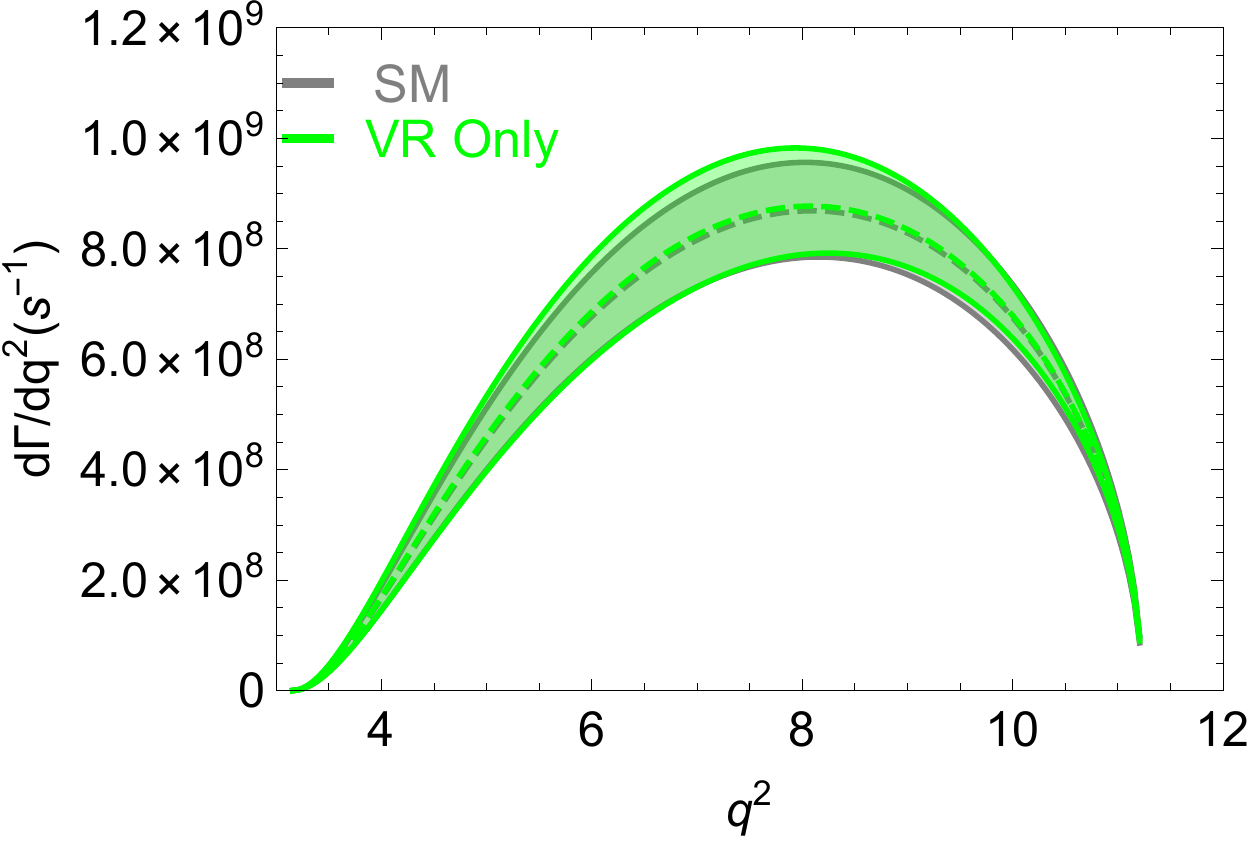}~
	\includegraphics[width=3.9cm,height=3.0cm]{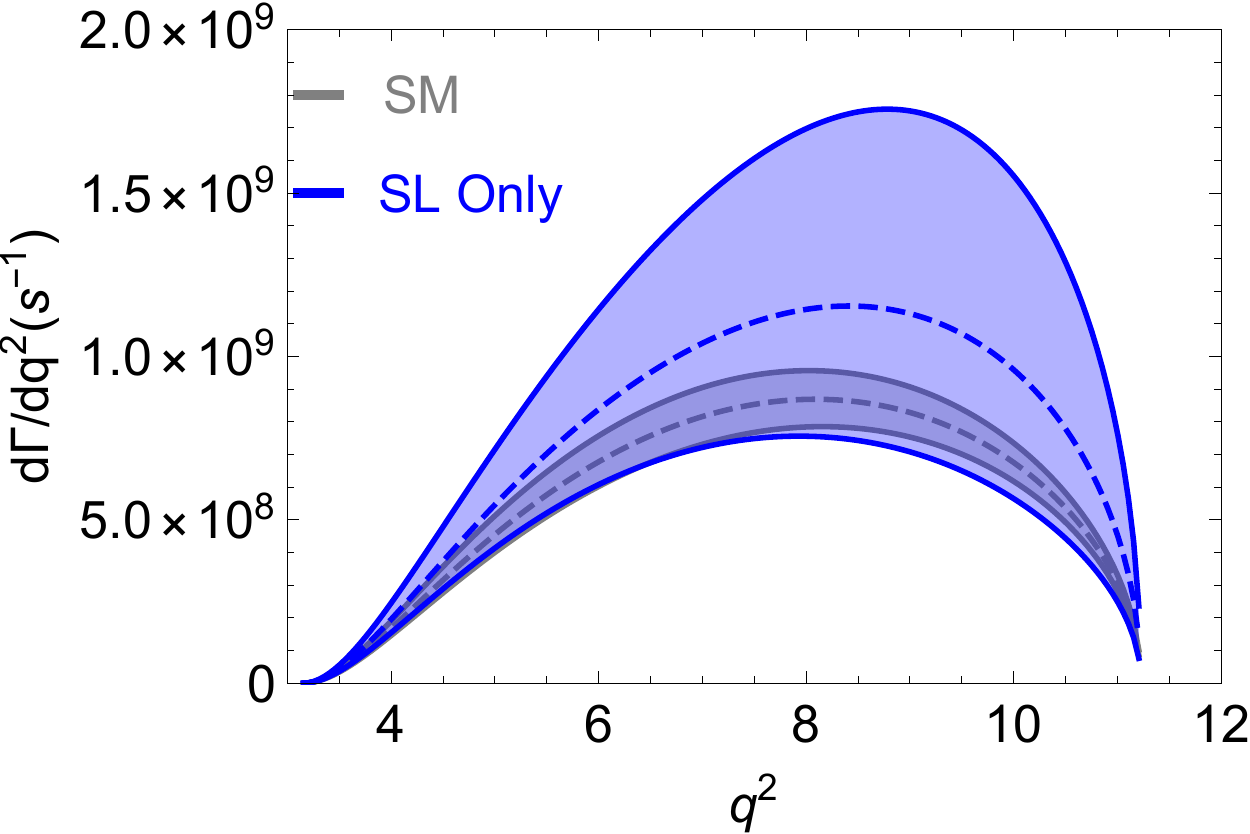}~
	\includegraphics[width=3.9cm,height=3.0cm]{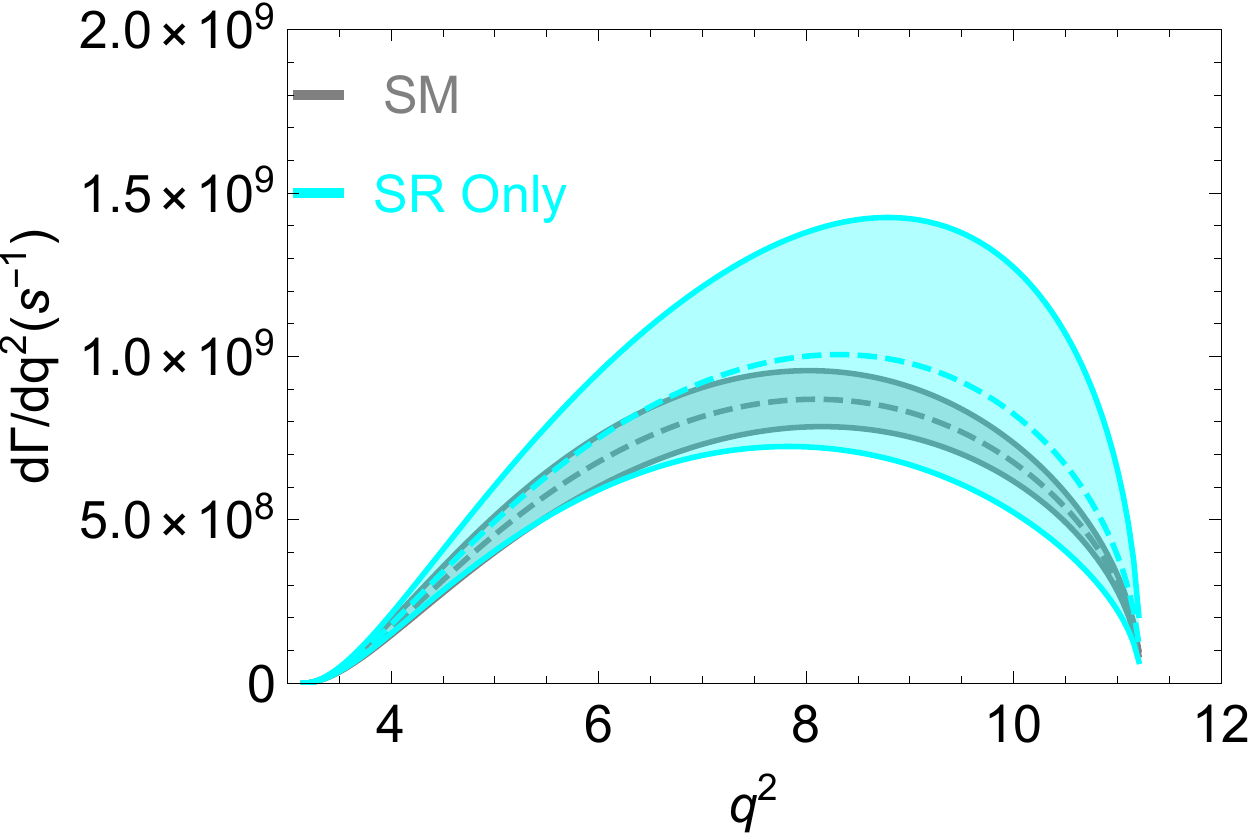}~\\
	\includegraphics[width=3.9cm,height=3.0cm]{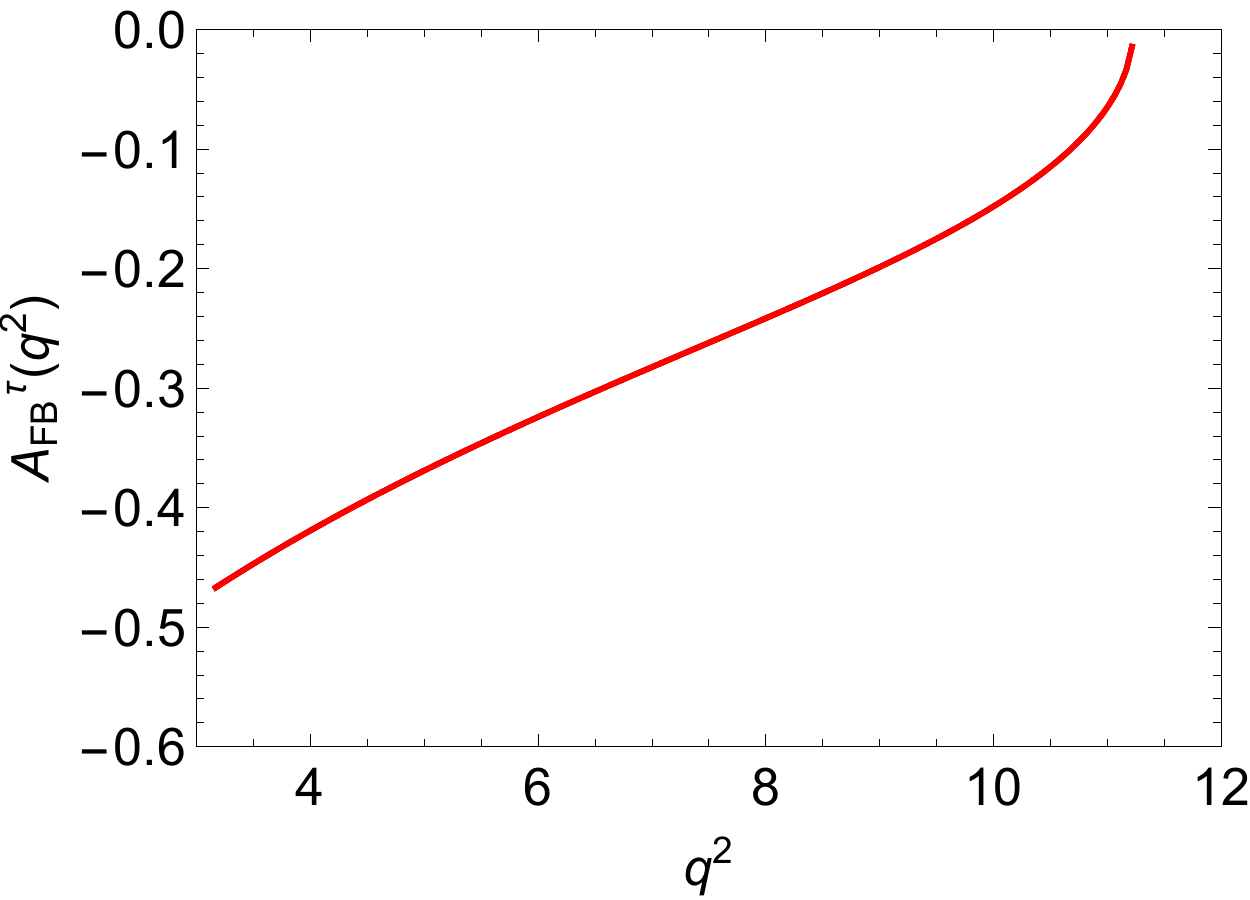}
	\includegraphics[width=3.9cm,height=3.0cm]{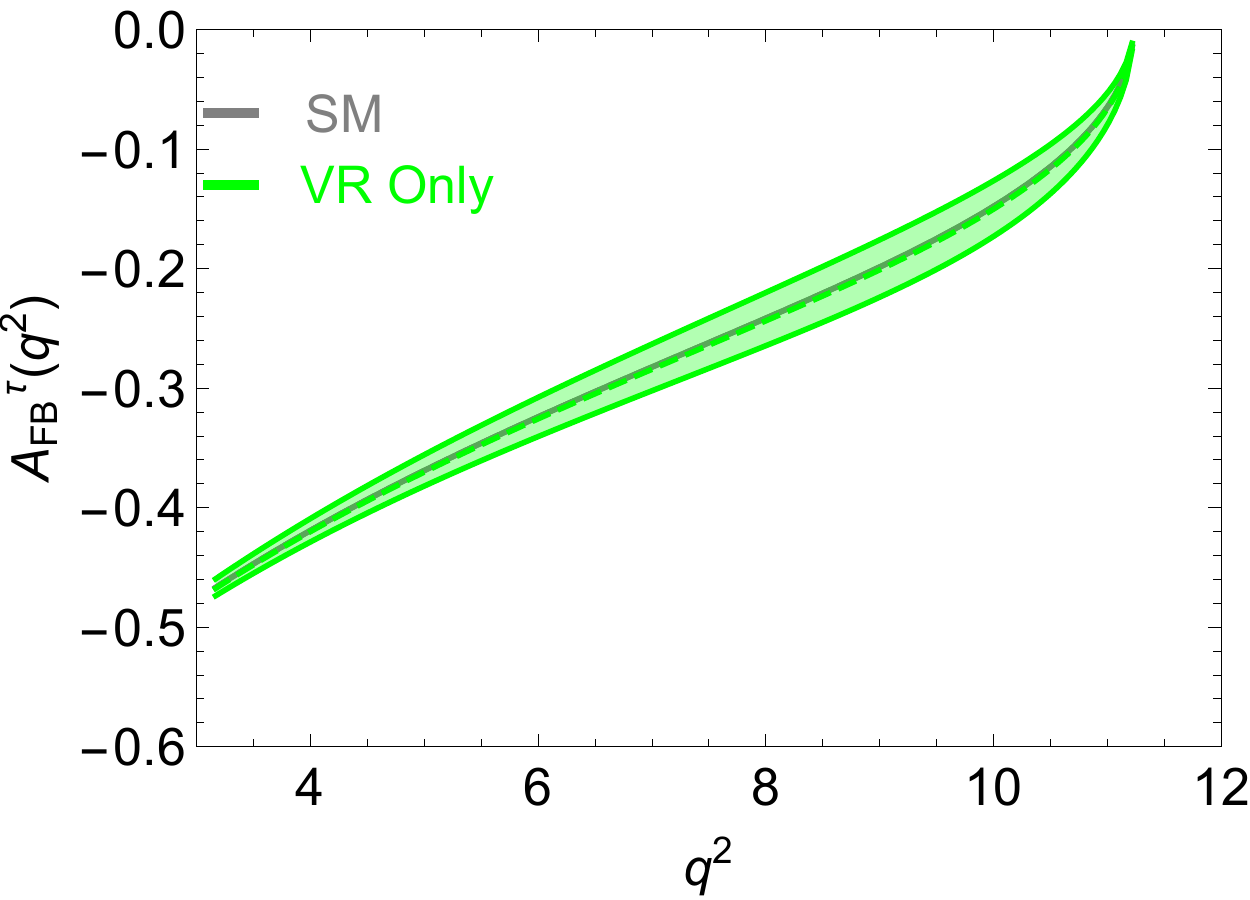}
	\includegraphics[width=3.9cm,height=3.0cm]{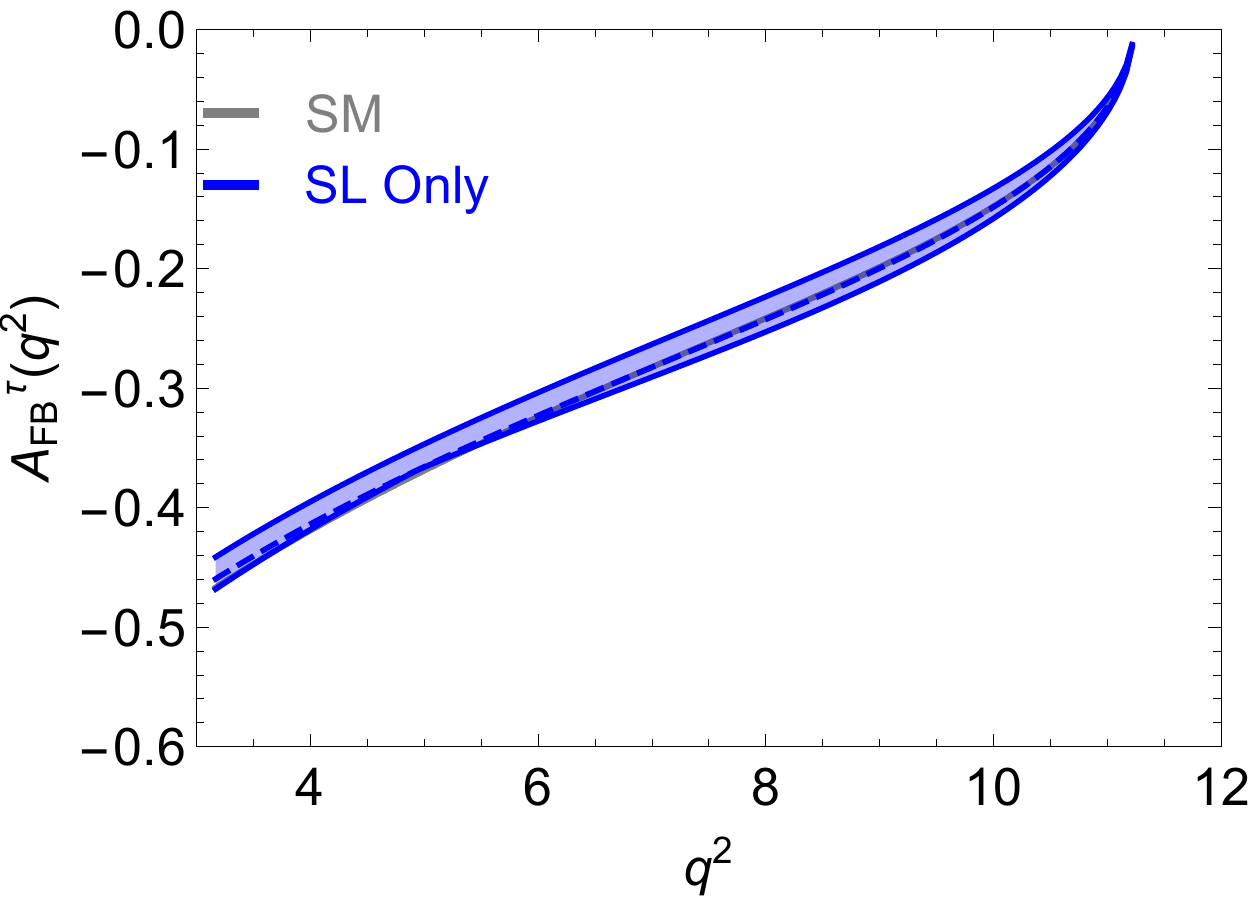}
	\includegraphics[width=3.9cm,height=3.0cm]{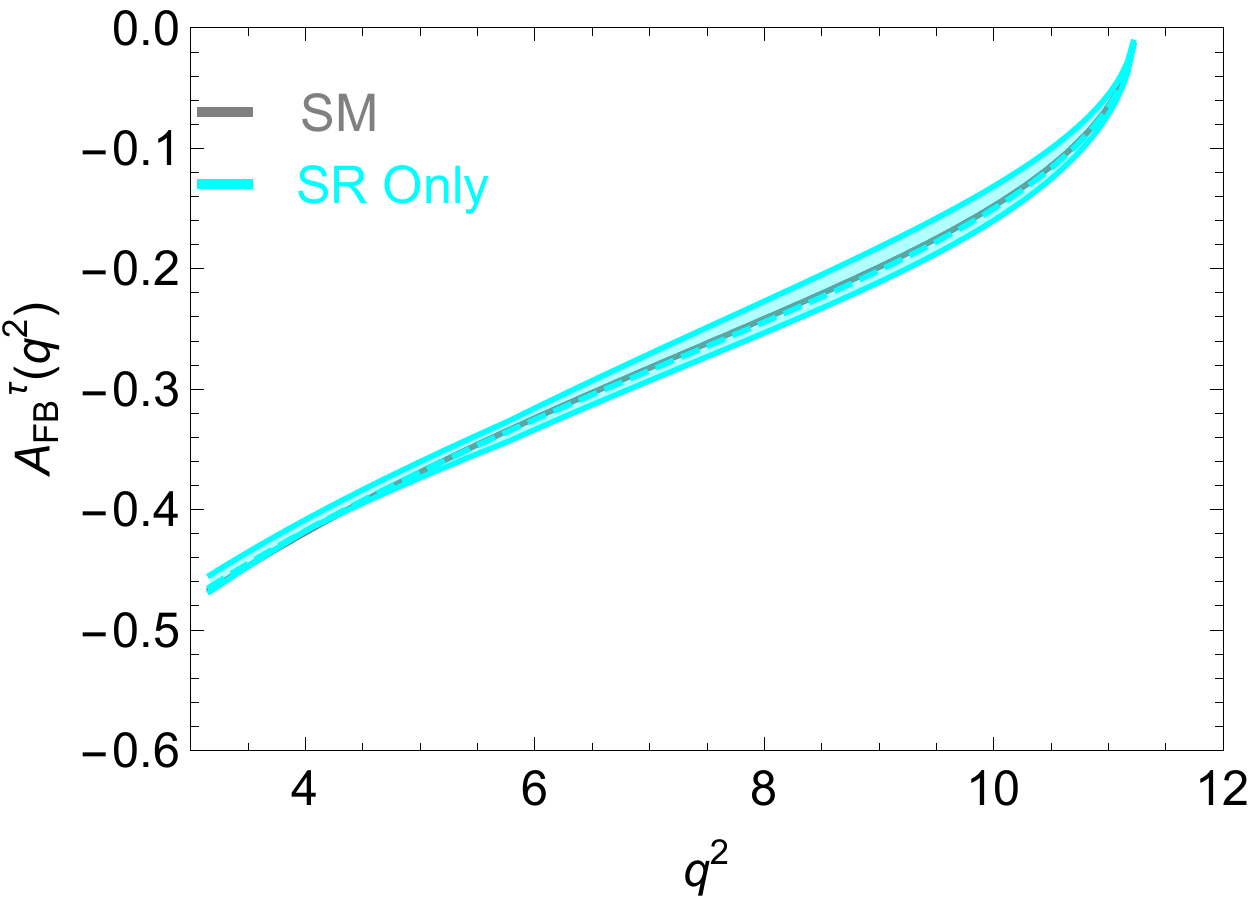}\\
	\includegraphics[width=3.9cm,height=3.0cm]{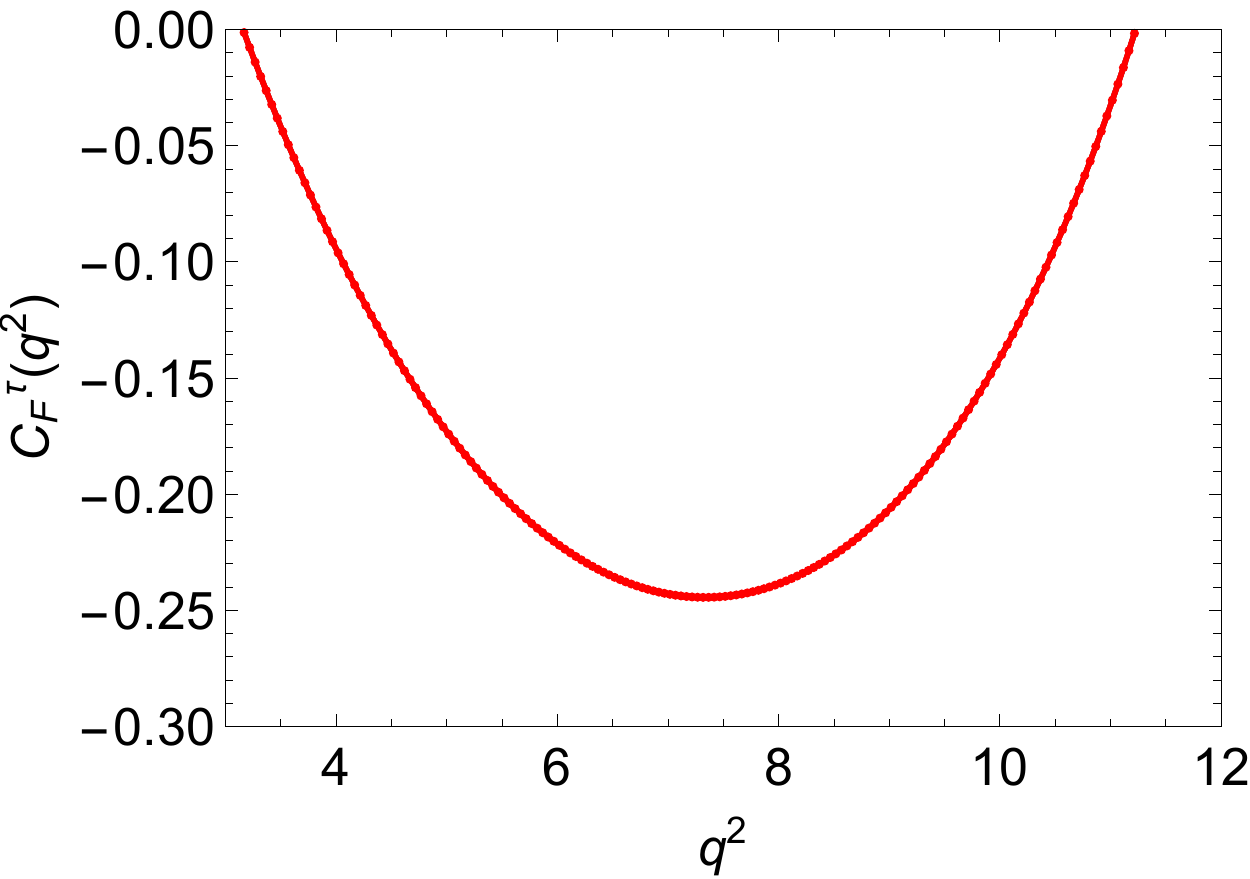}
	\includegraphics[width=3.9cm,height=3.0cm]{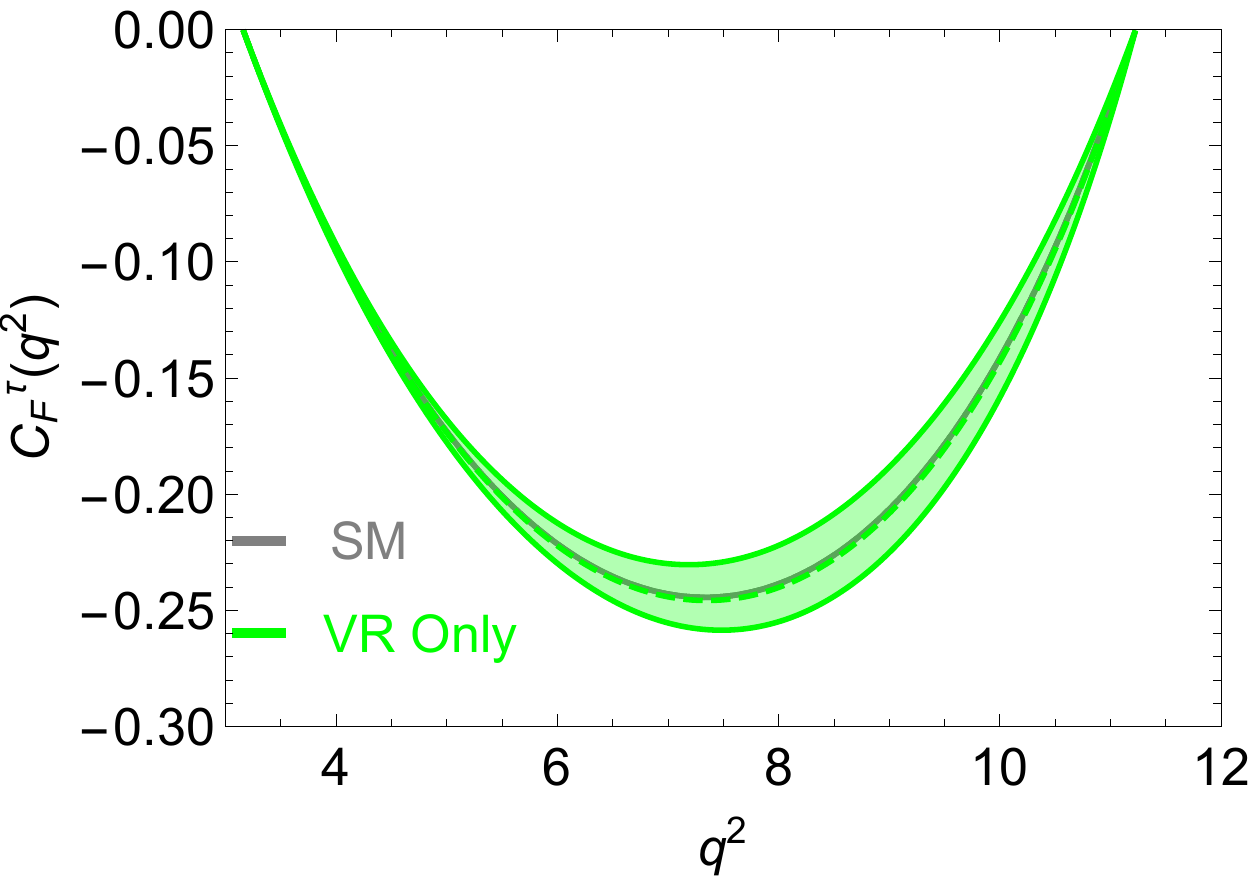}
	\includegraphics[width=3.9cm,height=3.0cm]{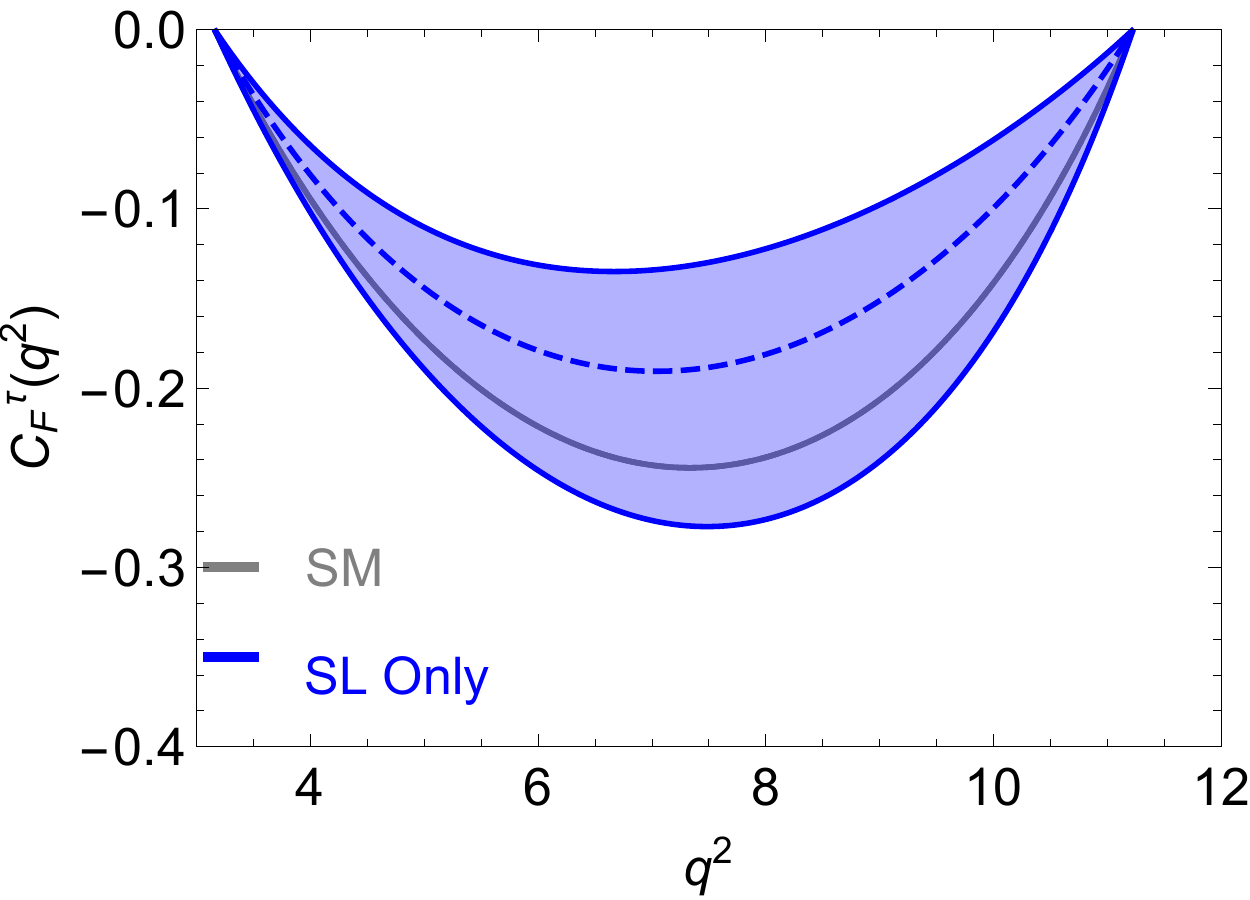}
	\includegraphics[width=3.9cm,height=3.0cm]{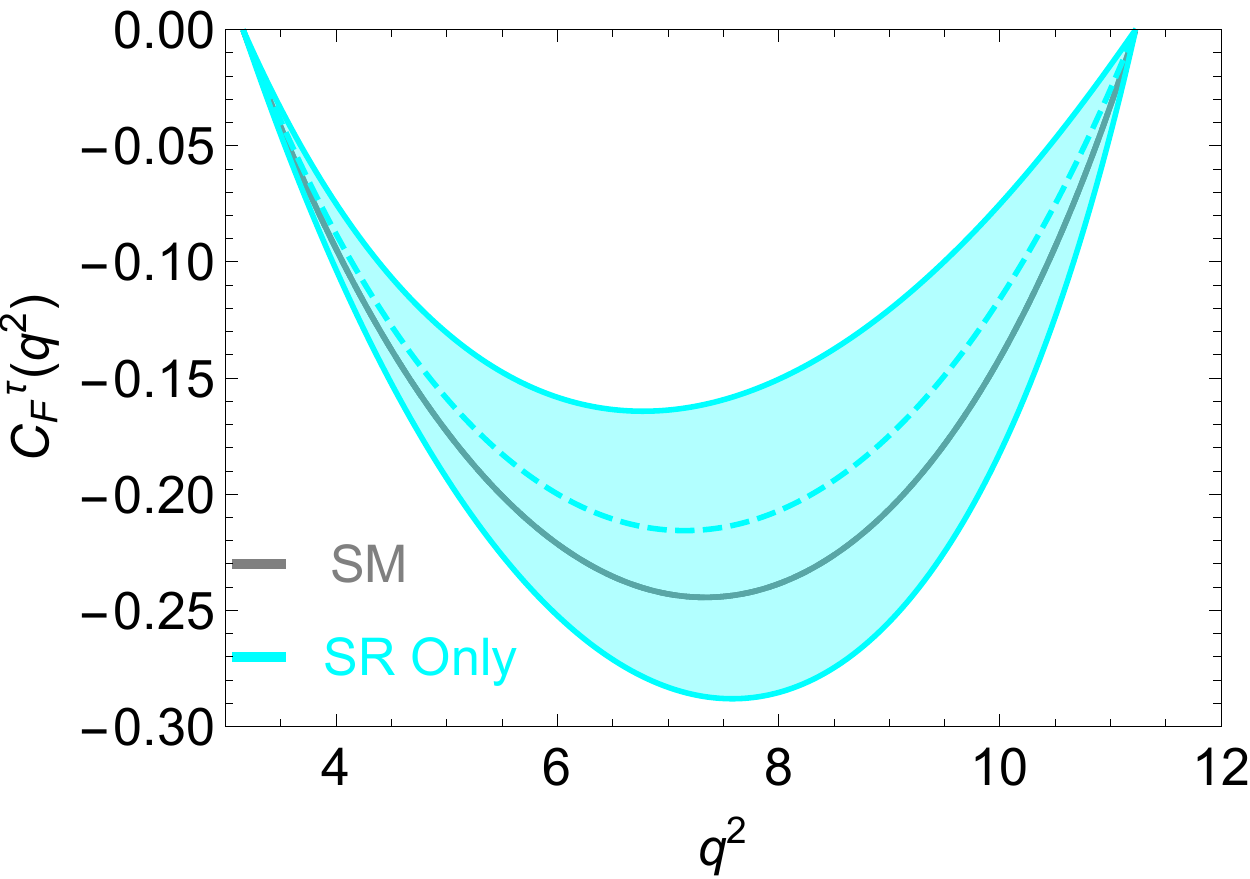}\\
	\includegraphics[width=3.9cm,height=3.0cm]{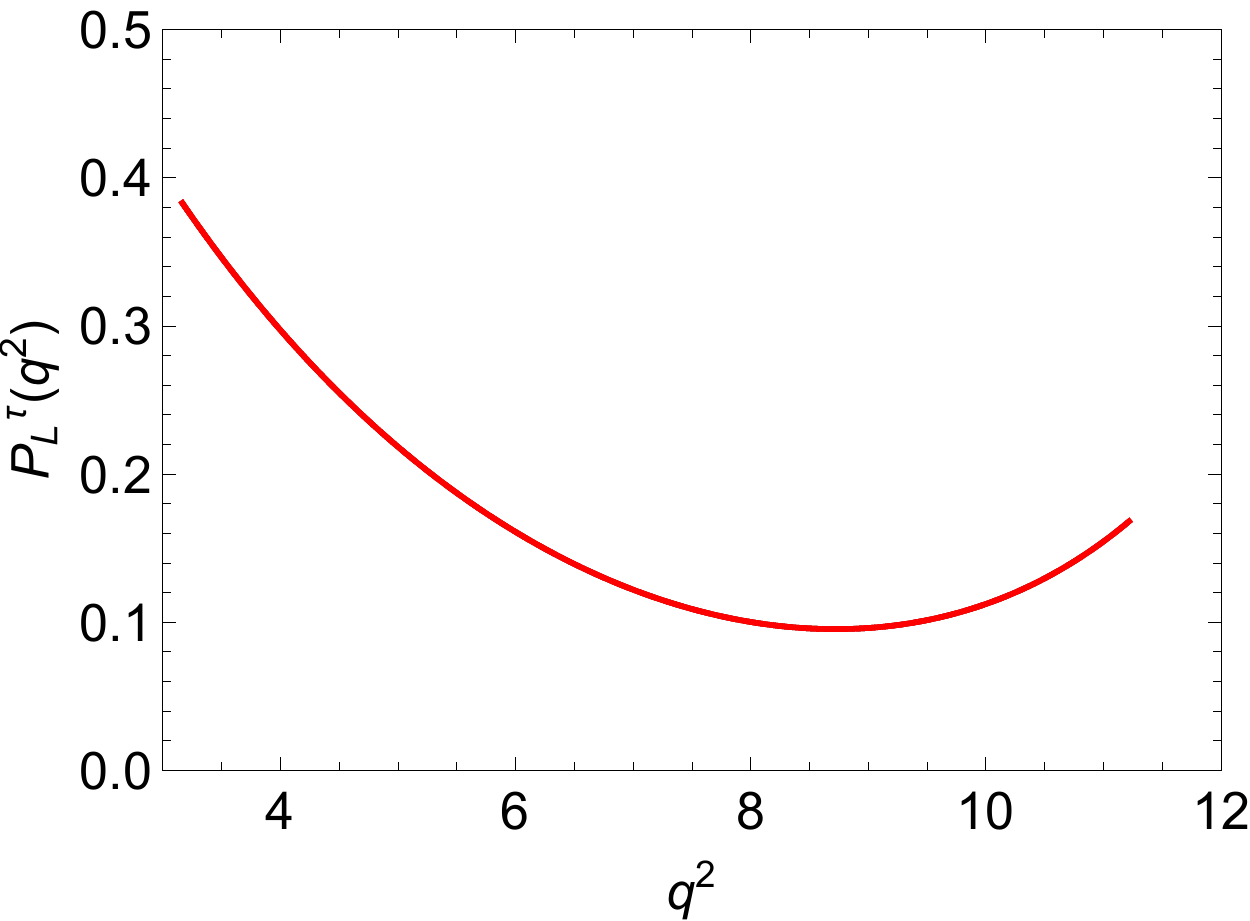}
	\includegraphics[width=3.9cm,height=3.0cm]{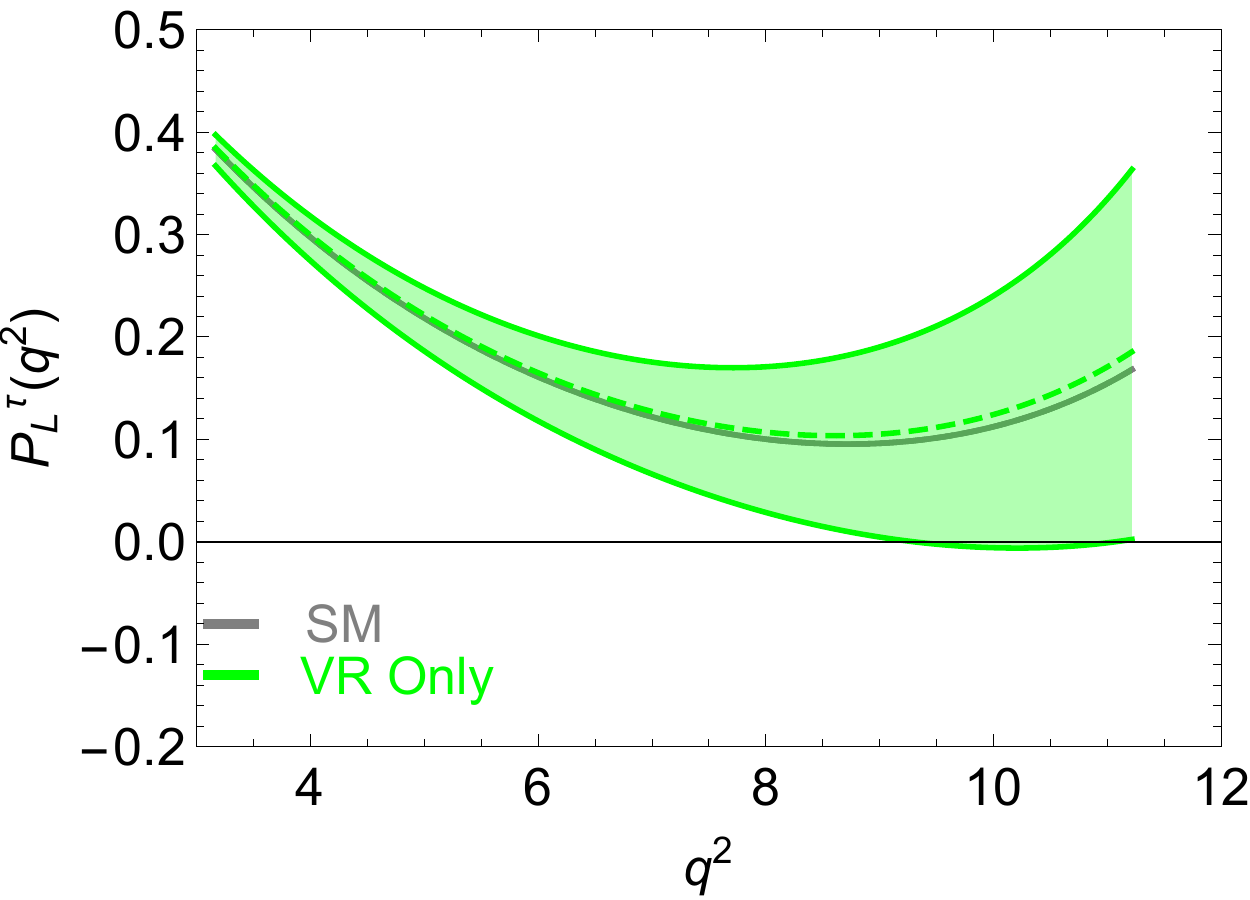}
	\includegraphics[width=3.9cm,height=3.0cm]{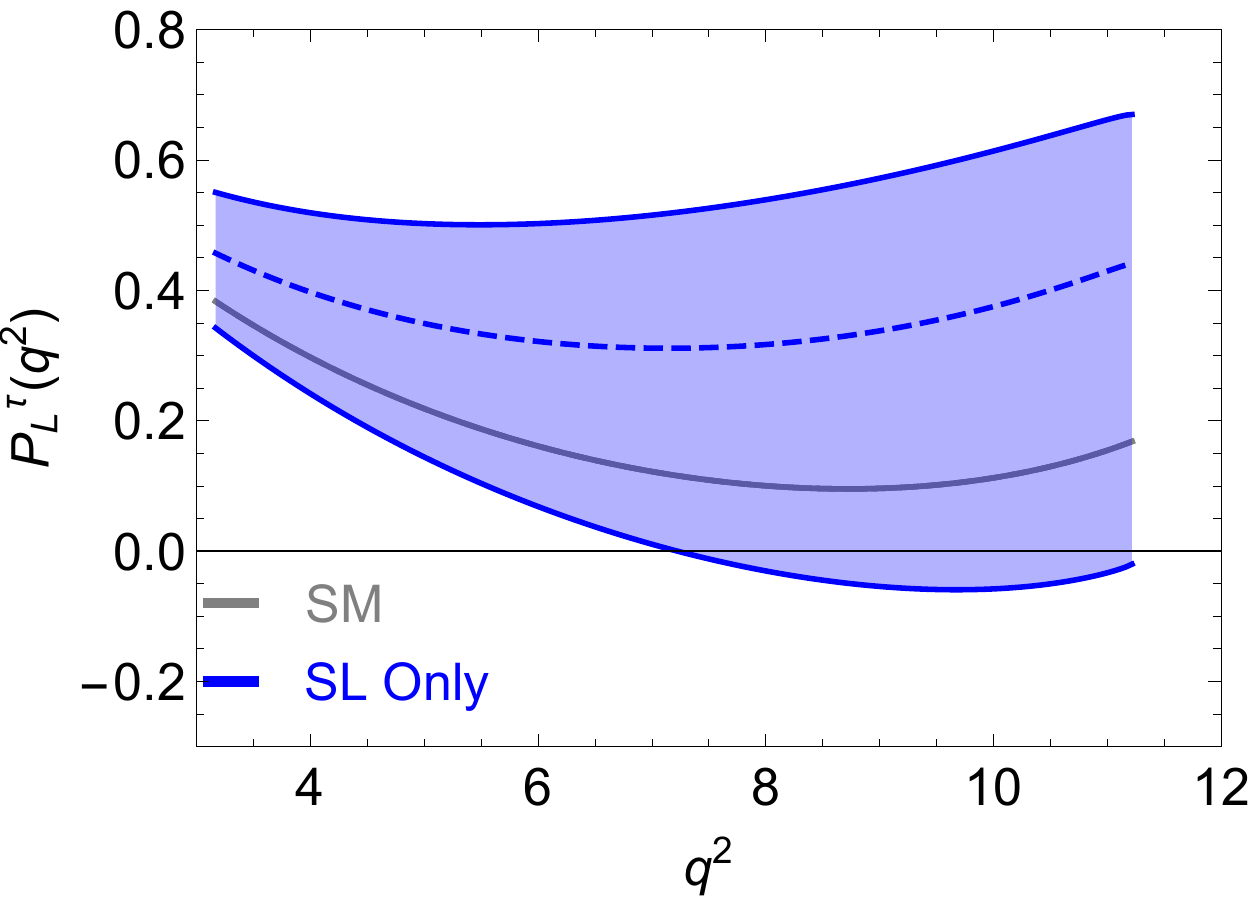}
	\includegraphics[width=3.9cm,height=3.0cm]{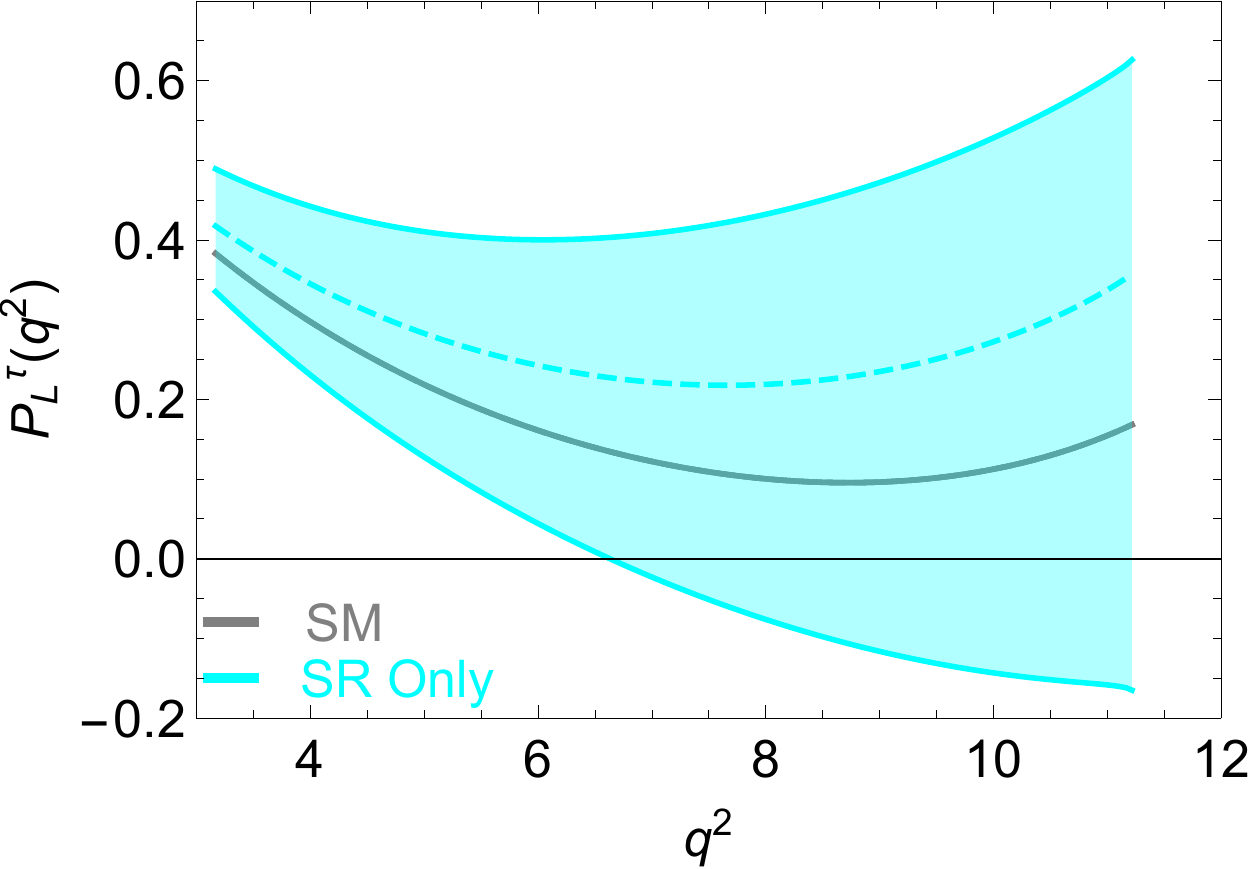}\\
	\includegraphics[width=3.9cm,height=3.0cm]{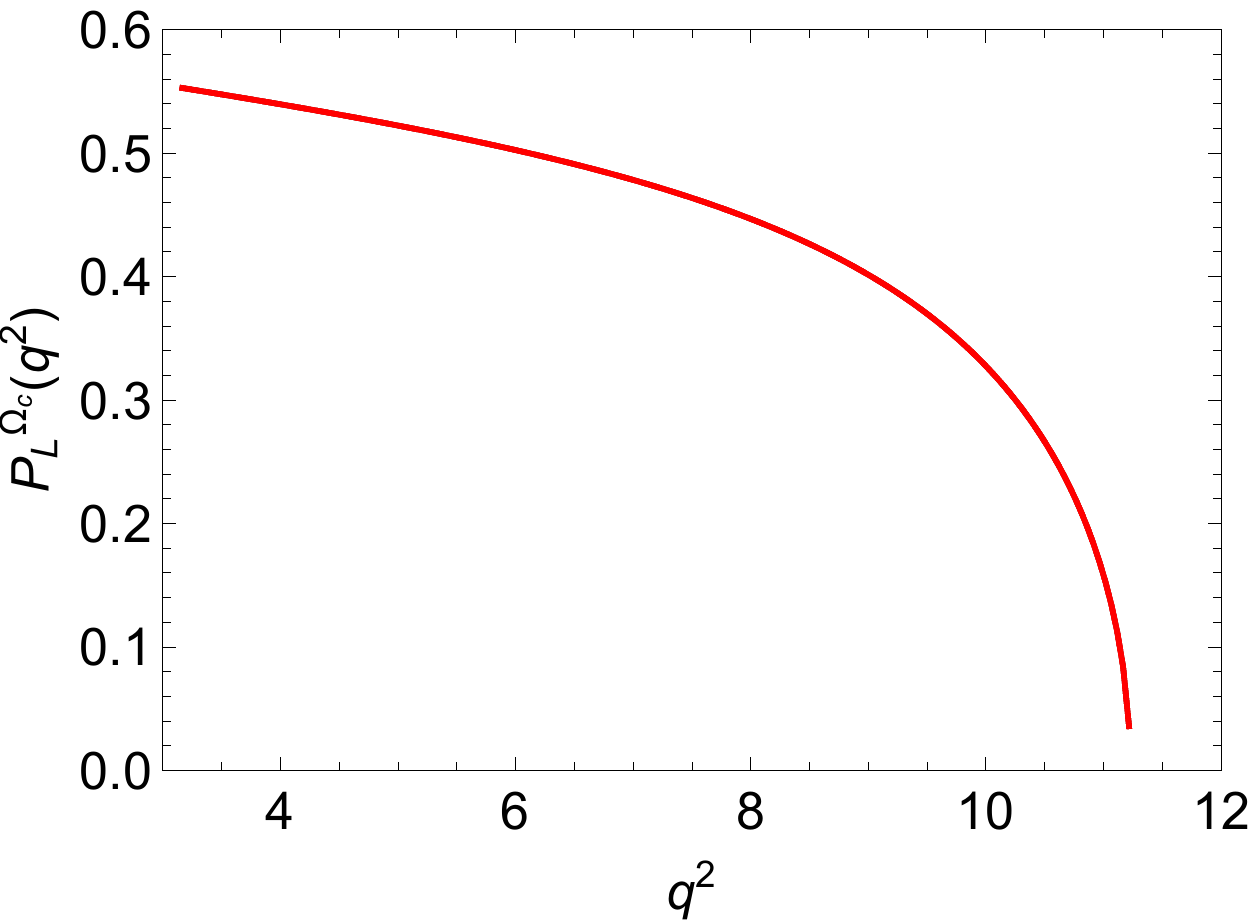}
	\includegraphics[width=3.9cm,height=3.0cm]{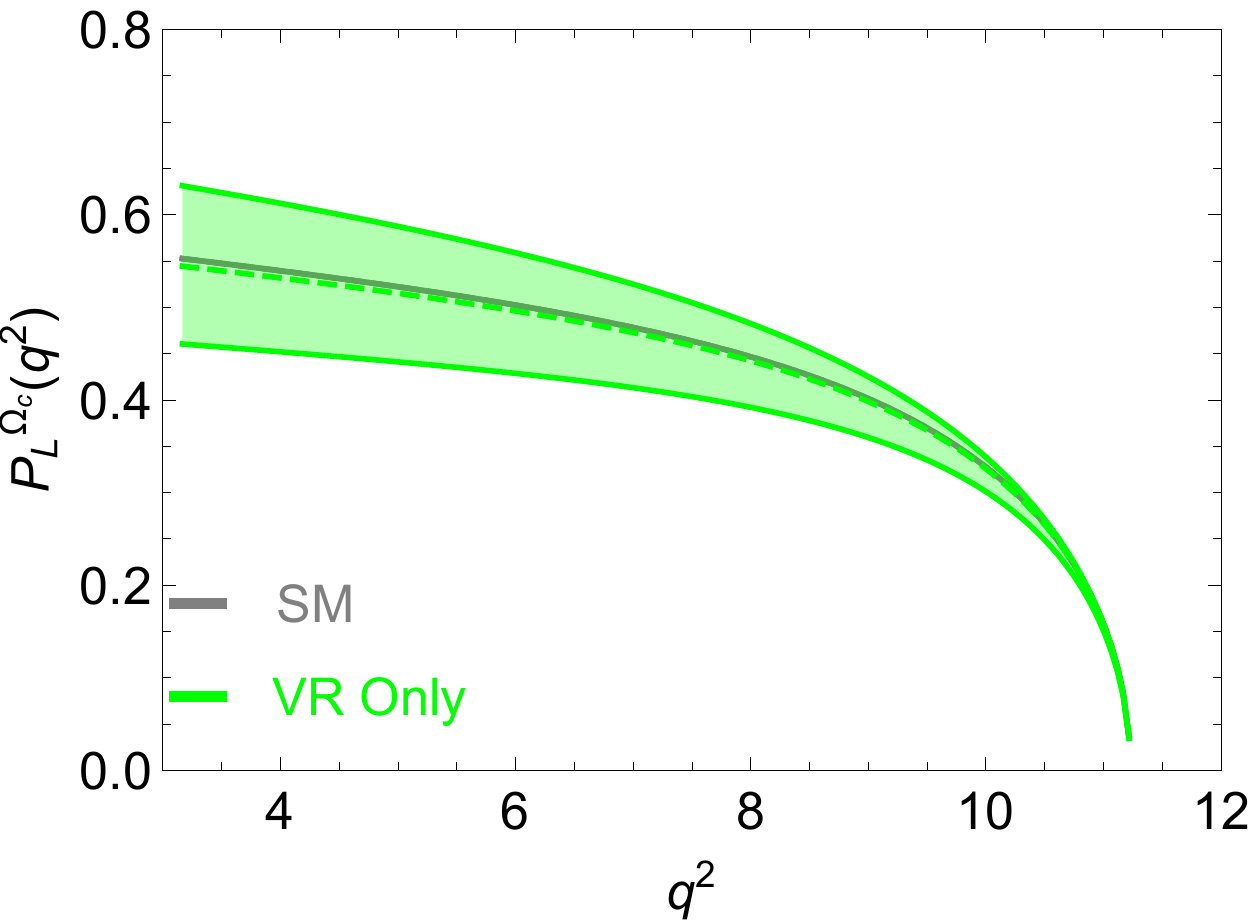}
	\includegraphics[width=3.9cm,height=3.0cm]{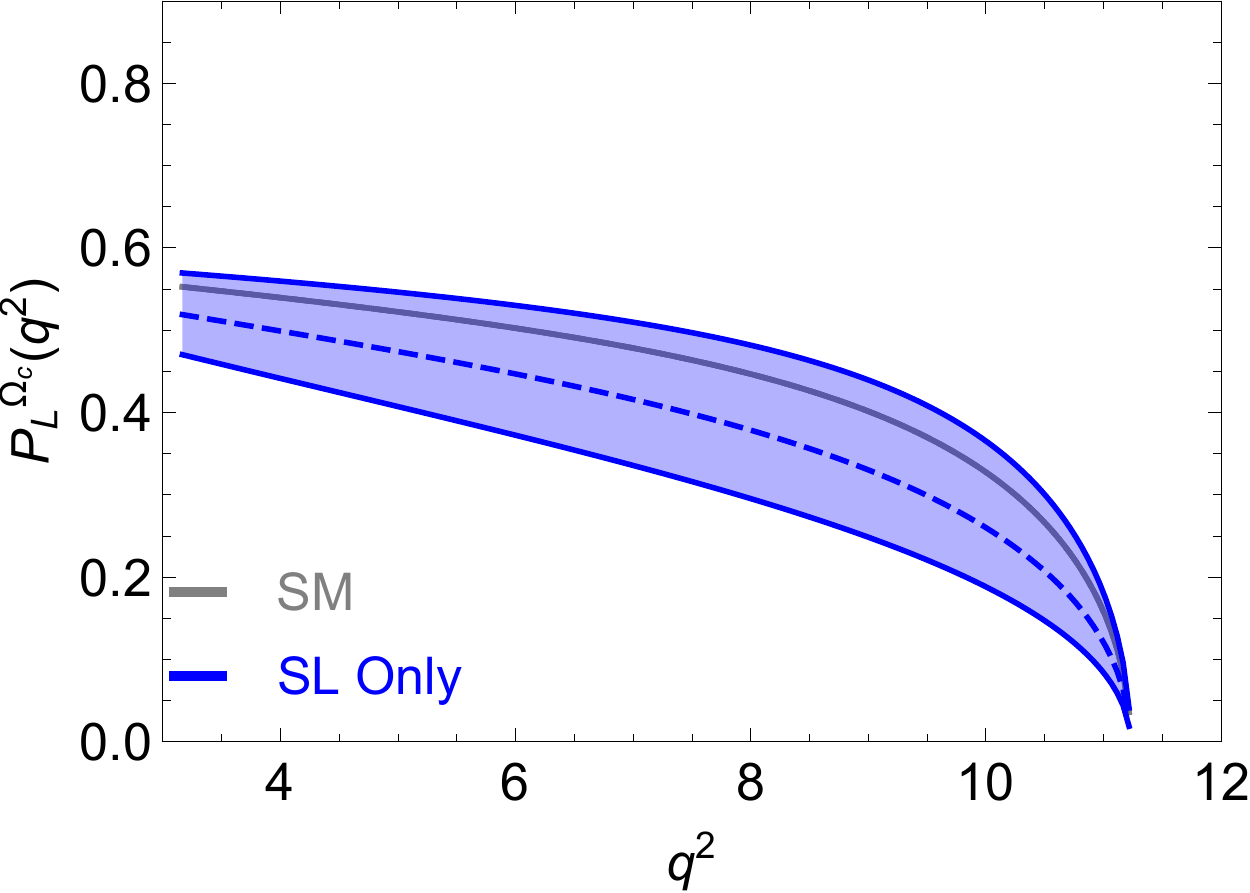}
	\includegraphics[width=3.9cm,height=3.0cm]{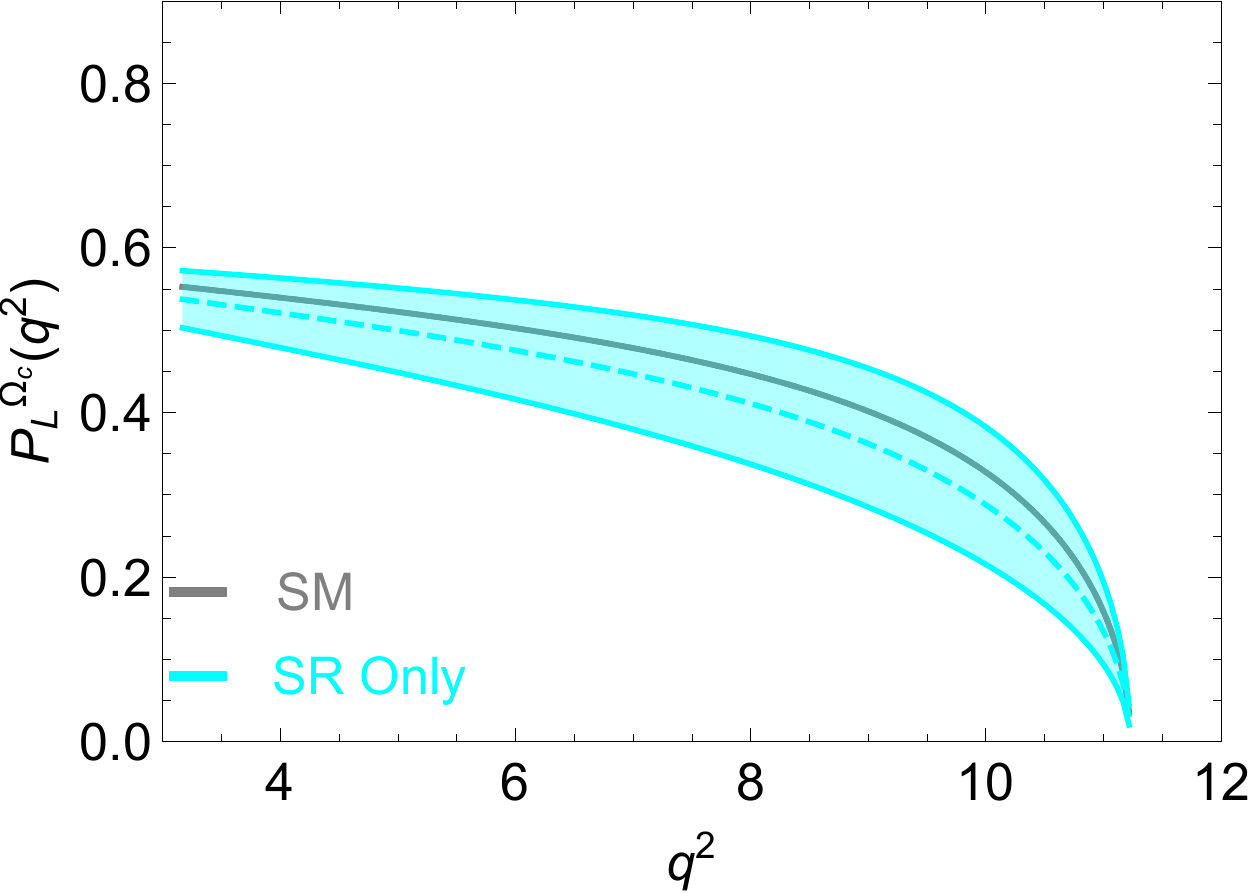}\\
	\includegraphics[width=3.9cm,height=3.0cm]{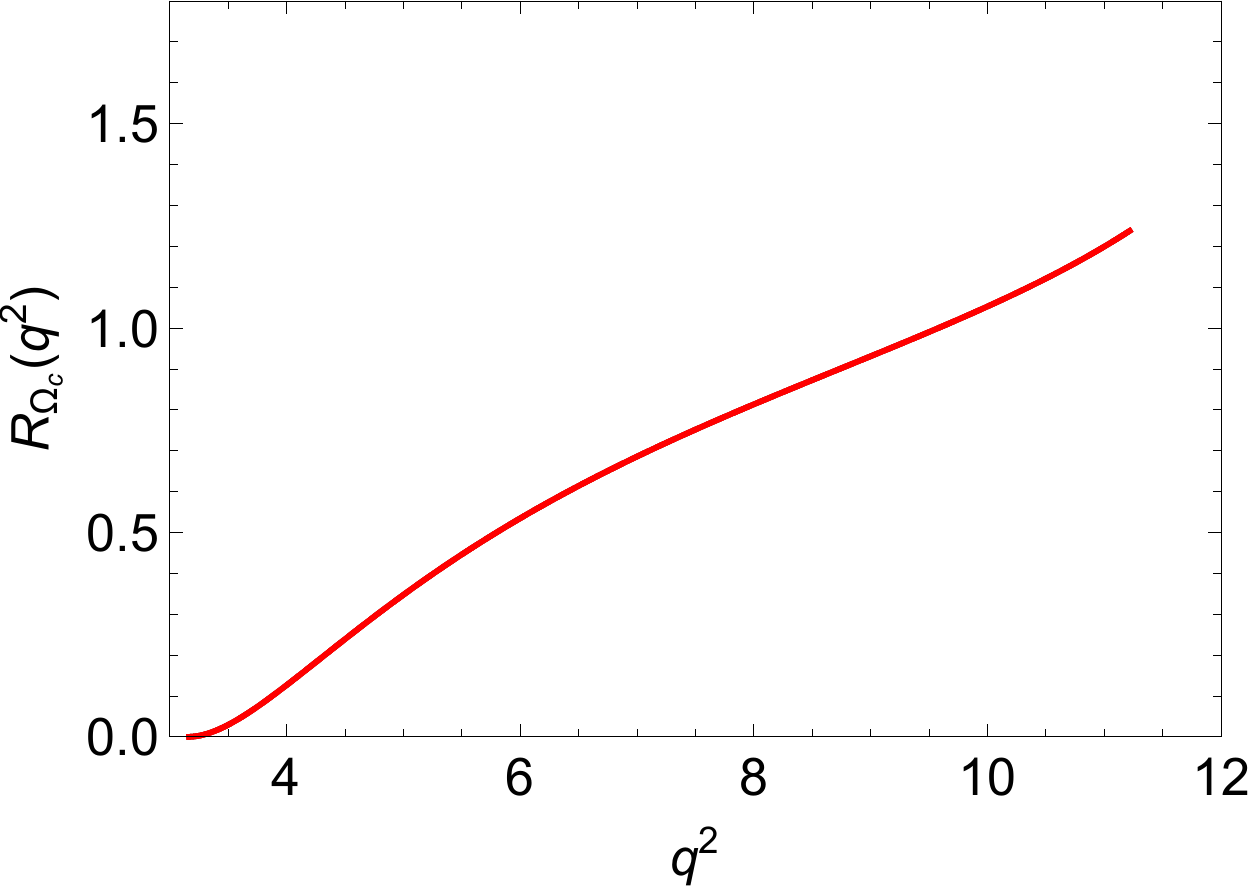}
	\includegraphics[width=3.9cm,height=3.0cm]{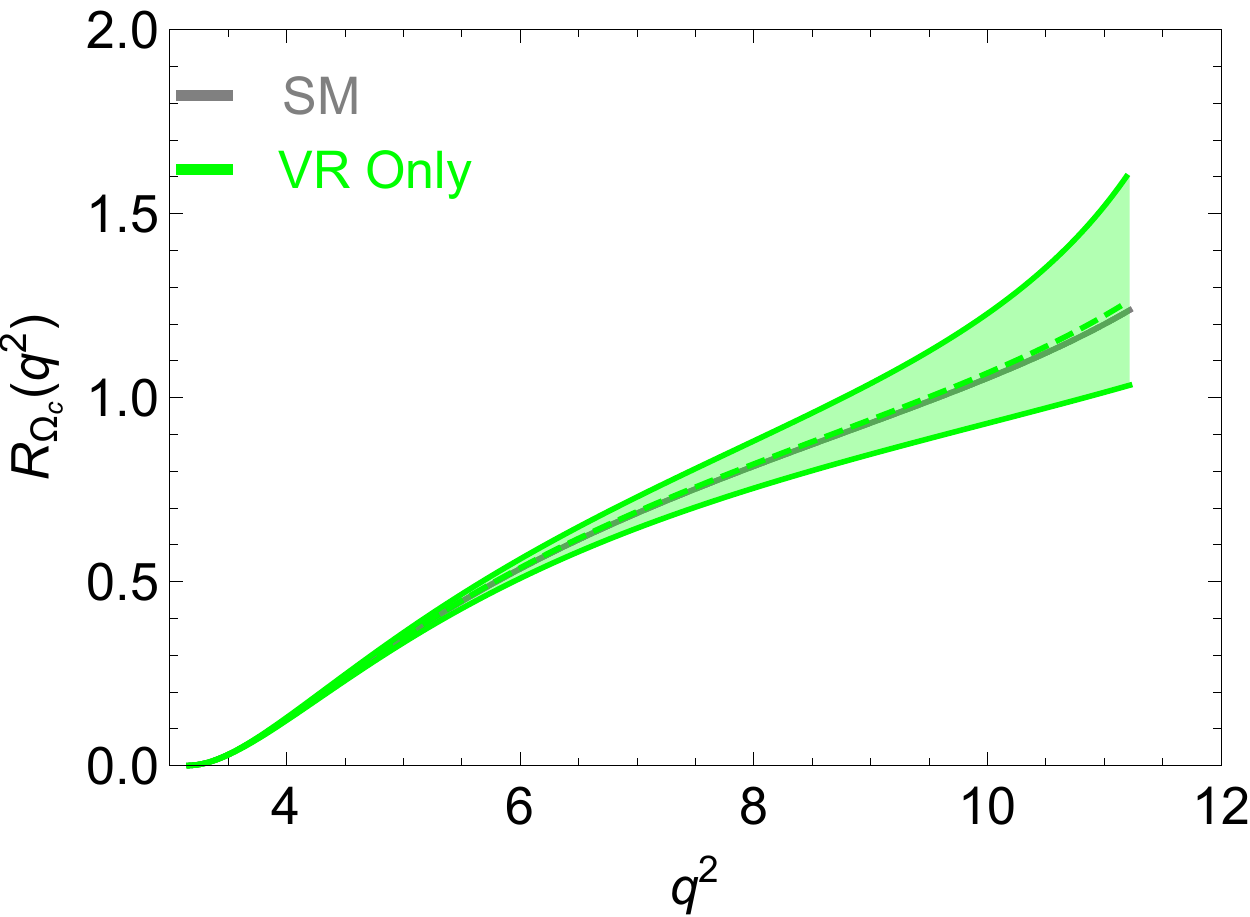}	
    \includegraphics[width=3.9cm,height=3.0cm]{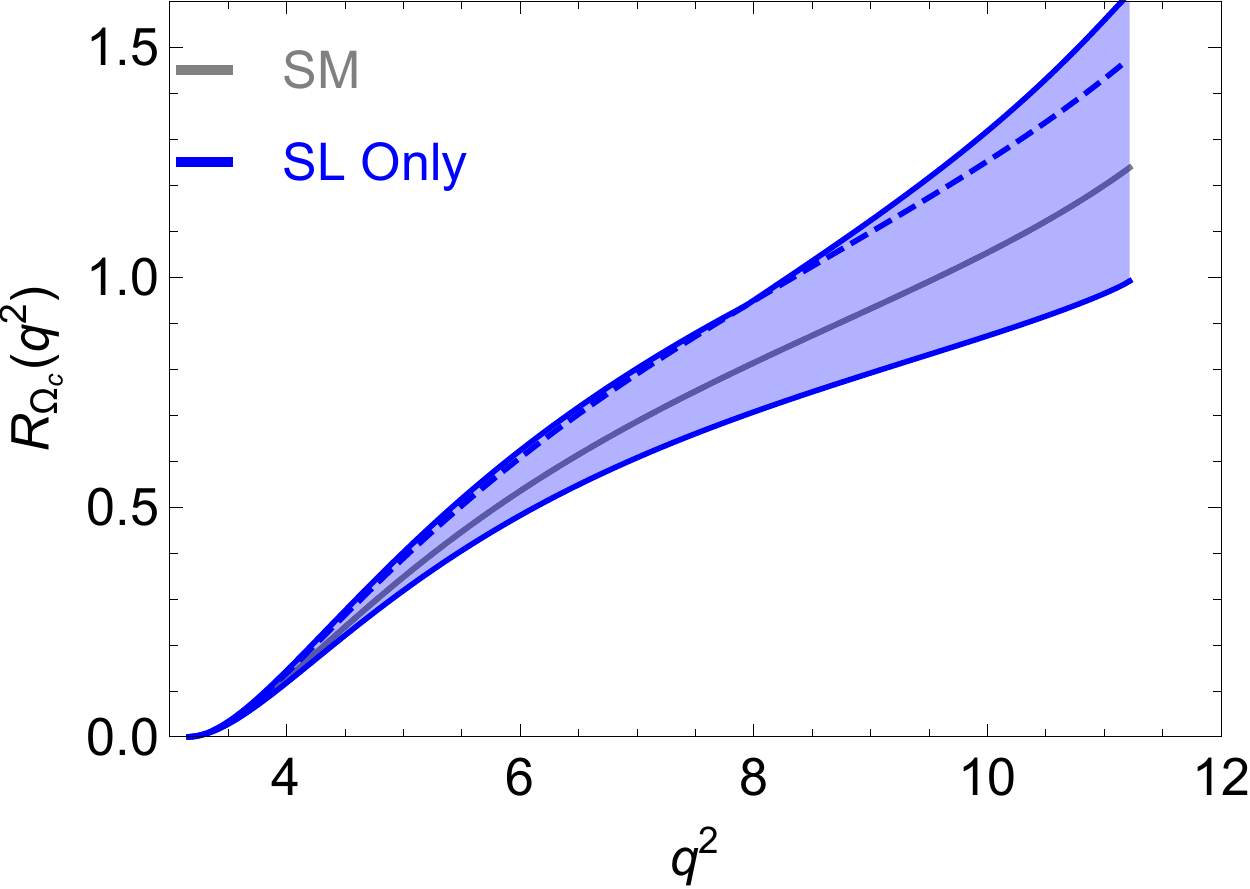}	
   \includegraphics[width=3.9cm,height=3.0cm]{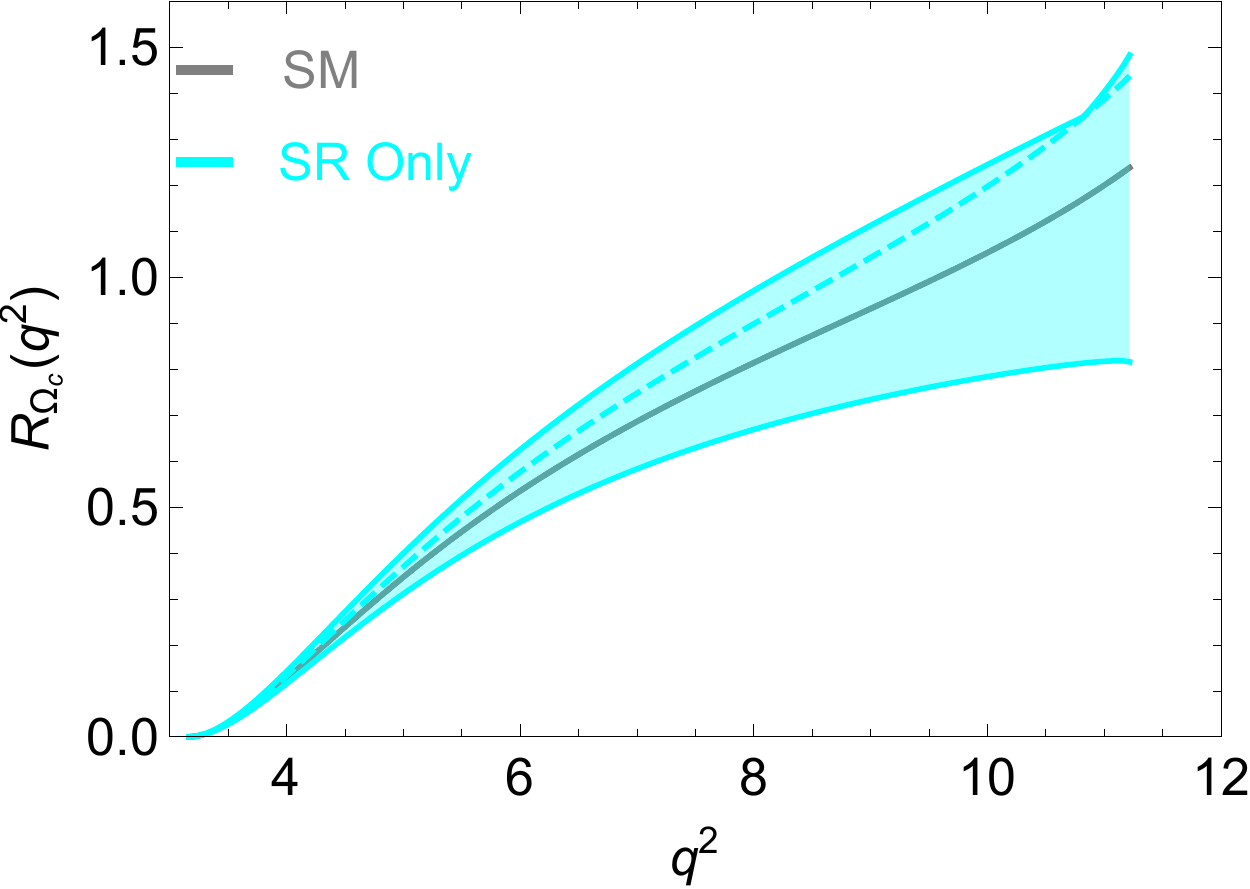}
	\caption{  {\small    The SM (gray) and NP  predictions in the presence of  $V_L$ (first column), $V_R$ (second column),
$S_L$ (third column) and $S_R$ (fourth column) coupling for the $q^2$ dependency observables
$d\Gamma/dq^2$, $A_{FB}^{\tau}(q^2) $, $C_F^{\tau}(q^2)$,  $P_L^{\tau}(q^2)$, $P_L^{\Omega_c}(q^2)$and ${\rm R}_{\Omega_c}(q^2)$
 relative to the decay $\Omega_b\to \Omega_c \tau \bar{\nu_{\tau}}$.
	}}\label{fig.ObNP1}
\end{figure*}

Next, we proceed to investigate the effects of these four NP coupling parameters $V_L$, $V_R$, $S_L$ and $S_R$  on the above observables
for various NP scenarios in a model independent way.
In order to avoid repetition, we only display the $q^2$ dependency of each observable for decay $\Omega_b \to \Omega_c \tau \bar{\nu}_{\tau}$
 and the results are displayed  in Fig.~\ref{fig.ObNP1}. 
In the figure we report the $q^2$ dependency of the observables $d\Gamma/dq^2$, $A_{FB}^{\tau}(q^2)$, $C_F^{\tau}(q^2)$, $P_L^{\tau}(q^2)$, $P_L^{\Omega_c}(q^2)$ and ${\rm R}_{\Omega_c}(q^2)$ for
$\Omega_b \to \Omega_c \tau \bar{\nu}_{\tau}$ transition including the contribution of only one NP vector or scalar type coupling parameter,
and we incorporate both SM and NP result.
In the Fig.~\ref{fig.ObNP1},
the band for the input parameters (form factors and $V_{cb}$) and different NP coupling parameters
restricted by the relative experimental  values of the processes $B\to D^{(*)} l \bar{\nu}_l$
and $B_c \to J/\psi l\bar{\nu}_l$ are represented with that different colors.
The SM and four NP scenarios are distinguished by gray (SM), red ($V_L$), green ($V_R$) , blue ($S_L$) and cyan ($S_R$)  colors, respectively.
In the Fig.~\ref{fig.ObNP1},  we suppose that the NP contributions only come from one NP coupling and we find the following remarks:

\begin{itemize}
	\item[*]
	When we only consider the effect of vector NP coupling $V_L$,
	the effect of this NP coupling appears in the $H^V_{\lambda_2,\lambda_W}$ and $H^A_{\lambda_2,\lambda_W}$ only.
   From Eq.~(\ref{Eq.dTds}), it is clear to find that the $d\Gamma/dq^2$ depends on  $(1+V_L)^2$ only.
	Using the constrained range of $V_L$ which are displayed in the  Fig.~\ref{fig.limitNP}, one can see
    that the deviation from the SM prediction due to the $V_L$
    coupling is observed only in the total differential decay rate $d\Gamma/dq^2$ and the observable is proportional to $(1+V_L)^2$.
	The $d\Gamma/dq^2$ is largely enhanced in the whole $q^2$ region.
	Moreover, the factor $(1+V_L)^2$ appears both in the numerator and denominator of the expressions which describe other observables
    simultaneously.
	So the NP dependency cancels in the ratios and we do not see any deviation from the SM prediction for other observables.

	\item[*]  Similar to $V_L$, the  NP coupling  parameter $V_R$ is also included in the vector and the axial-vector helicity amplitudes.
	In this case, the $d\Gamma/dq^2$ depends on both $(1+V_R)^2$ and $(1-V_R)^2$. Hence, there is no cancellation of NP
    effects in the
    ratios and there is deviation in each observable from the SM prediction.
	The deviation of  $d\Gamma/dq^2$ from their SM prediction is not so significant,
    while, it  is very significant for other observables.
	The effects of the $V_R$ coupling are rather significant on the observables $P_L^{\tau}(q^2)$, $P_L^{\Omega_c}(q^2)$
   and ${\rm R}_{\Omega_c}(q^2)$, especially in largest $q^2$ region for $P_L^{\tau}(q^2)$ and ${\rm R}_{\Omega_c}(q^2)$ and lowest $q^2$ region for $P_L^{\Omega_c}(q^2)$.
	

	\item[*]
    The effects of the scalar NP coupling $S_L$ come into the scalar and pseudoscalar helicity amplitudes $H_{\lambda_2,\lambda_W}^S$ and $H_{\lambda_2,\lambda_W}^P$.
    One can see that it is  different from $V_L$ and $V_R$ coupling scenarios.
    From Eq.~(\ref{Eq.dTds}) one can see that $d\Gamma/dq^2$ depends on $S_L$ and  $S_L^2$ in this case.
	So there is also no cancellation in the numerator and denominator of the expressions in other observables simultaneously.
	We can find that the deviation from their SM prediction is more pronounced than that with $V_L$ and $V_R$ NP coupling except
    $A_{FB}^{\tau}(q^2)$.
	The deviation from the SM prediction for $d\Gamma/dq^2$ is most prominent at
	$q^2 \approx 8.8GeV^2 $.
	When consider the value of the $S_L$ NP coupling, there may or may not be a zero crossing in the  $P_L^{\tau}(q^2)$, while there is
    no zero crossing for $P_L^{\tau}(q^2)$ in the SM prediction.
	Besides, the deviations from their SM prediction for $P_L^{\tau}(q^2)$ and ${\rm R}_{\Omega_c}(q^2)$ are most prominent at largest $q^2$
    region.
	There are some differences between our results and Ref.~\cite{Rajeev:2019ktp} for $S_L$  NP coupling scenario.
    In Ref.~\cite{Rajeev:2019ktp}, there are two constraint results for $S_L$ NP coupling and they are $S_L \in [-0.2,0.1]$ and $[-1.6,-1.4]$ respectively.
    The authors use $S_L \in [-1.6,-1.4]$ when consider the NP effect of $S_L$.
    If $S_L \in [-0.2,0.1]$ in their analysis, their result are similar to our work for this scenario.

	\item[*]
    From last column in Fig.~\ref{fig.ObNP1} considering the $S_R$ NP coupling, the change trend of each observable are similar to the $S_L$ scenario.
    Because NP effects which come from the $S_R$  NP coupling are also encoded in the scalar and pseudoscalar helicity amplitudes only,
    the $d\Gamma/dq^2$ depends on $S_R$ and $S_R^2$.
	The deviation from the SM prediction of  $d\Gamma/dq^2$ may be less obvious than the  $S_L$ scenario. However,  it is larger for
     $C_F^{\tau}(q^2)$ compared to $S_L$ scenario.
	In the $P_L^{\tau}(q^2)$, the zero -crossing point may shift slightly towards a lower $q^2$ value than in the $S_L$ case.
	
\end{itemize}


Finally, we also explore the  impact of  these four combinations for vector and scalar type couplings such as
$V_L$-$S_L$, $V_L$-$S_R$, $V_R$-$S_L$, and $V_R$-$S_R$ to above various observables for $\Omega_b \to \Omega_c \tau \bar{\nu}_{\tau}$ process.
We find that the NP predictions of the same observables in these four combinations NP scenarios  show a similar variation tendency to the increasing of $q^2$ and have similar deviations to their corresponding $S_L$ and $S_R$ predictions,
except that the value of the corresponding longitudinal axis is different.
In order to avoid repetition, we do not display the results of different combinations anymore.
At the same time we find that similar conclusions can be also  made for the $\Sigma_b\to \Sigma_c \tau\bar{\nu}_{\tau}$ decay process.
\section{Summary}\label{Sec.four}
Several anomalies ${\rm R}_{D^{(*)}}$ and ${\rm R}_{J/\psi}$ observed in the semileptonic B meson decays have indicated the hints of  LFUV and attracted the attention of many researchers.
Many works about baryon decays  $\Lambda_b \to \Lambda_c(p)l \bar{\nu}_l$ and  $\Xi_b \to \Xi_c(\Lambda) l \bar{\nu}_l$ have been done to investigate the NP effects of above anomalies on the precess $b\to c(u) l \bar{\nu}_l$.
These baryon decays not only can provide an independent determination of the CKM matrix element $|V_{cb}|$
but also may be further confirmation of the hints of LFUV that is helpful in exploring NP.
At present, there exist few quantitative measurement for the semileptonic decay of $\Omega_b$ and $\Sigma_b$
due the complexity  baryons structures and the lack of precise predictions of various form factors.
It is indeed necessary to investigate the semileptonic baryon decays $\Omega_b\to \Omega_c l \bar{\nu}_l$ and $\Sigma_b\to \Sigma_c  l \bar{\nu}_l$ both theoretically and experimentally to test the LFUV.


In this work we have used the helicity formalism to get various angular decay distribution
 and have performed a model independent analysis of baryonic $\Omega_b\to \Omega_c l \bar{\nu}_l$ and $\Sigma_b\to \Sigma_c l \bar{\nu}_l$ decay processes.  In this work we considered the NP coupling parameters to be complex in our analysis.
In order to constrain the  various NP coupling parameters, we have assumed that only one NP coupling parameter is present one time.
We have gotten strong bounds on the phases and strengths of the various NP coupling parameters from the latest experimental limits of
$B \to D^{(*)} l \bar{\nu}_l $ and $B_c \to J/\psi l \bar{\nu}_{l}$.
Using the constrained NP coupling parameters, we  have estimated various observables of  the $\Omega_b\to \Omega_c l \bar{\nu}_l$ and $\Sigma_b\to \Sigma_c  l \bar{\nu}_l$ baryon decays in the SM and various NP scenarios in a model independent way.
The numerical results have been presented for $e$, $\mu$ and $\tau$ mode respectively in  SM.
We also display the $q^2$ dependency of different observables  for  $\Omega_b \to \Omega_c \tau \bar{\nu}_{\tau}$ process within the SM and various NP coupling scenarios.
The results show that $d\Gamma/dq^2$ including any kind of NP couplings  are all enhanced largely and have significant deviations
comparing to their SM predictions in whole $q^2$ region.
In the $V_L$ scenario, the observables $A_{FB}^{\tau}(q^2)$, $C_F^{\tau}(q^2)$, $P_L^{\tau}(q^2)$, $P_L^{\Omega_c(\Sigma_c)}(q^2)$ and ${\rm R}_{\Omega_c(\Sigma_c)}(q^2)$
are the same as their corresponding SM predictions  because the coefficient $(1+V_L)^2$ appears in the numerator and the denominator of the  expressions which describing these observables simultaneously.
We noticed a profound deviation in all angular observables of the semileptonic baryonic $b\to c \tau \bar{\nu}_{\tau}$ process due to the additional contribution of  $V_R$, $S_L$ and $S_R$ couplings to the SM.
The deviations from their SM prediction of $P_L^{\tau}(q^2)$ and ${\rm R}_{\Omega_c}(q^2)$  are most prominent at largest $q^2$ region.

Till now there are only some  experimental data about the non-leptonic  decay of  $\Omega_b$ and $\Sigma_b$,
and there is poor quantitative measurement of the semileptonic decay rates of $\Omega_b$ and $\Sigma_b$,
Though there is no experimental measurement on these baryonic $b\to c l \bar{\nu}_l$ decay processes, the study of this work is  found to be very crucial in order to shed light on the nature of NP.
In the near future, more data on $\Omega_b$ will be obtained by the LHCb experiments
and we hope the results of the  observables discussed in this work can be tested at experimental facilities at BEPCII, LHCb and Belle II.

\section*{Acknowledgements}
We would like to thank  Yuan-Guo Xu for providing us some  helpful discussion and constant encouragement on the manuscript.
 This work was supported by the National Natural Science Foundation
of China (Contracts Nos.~11675137 and 11947083) and the Key Scientific Research Projects of Colleges and Universities in Henan Province (Contract No. 18A140029).

\end{document}